\RequirePackage{ifpdf}
\ifpdf 
\documentclass[pdftex]{sigma}
\else
\documentclass{sigma}
\fi

\usepackage{slashed}

\newcommand\sfrac[2]{{\textstyle\frac{#1}{#2}}}
\newcommand\ignore[1]{}
\def\one{{\,\hbox{1\kern-.8mm l}}}

\newcommand{\SO}{\mathrm{SO}} 
\newcommand{\SU}{\mathrm{SU}} \newcommand{\U}{\mathrm{U}}
\newcommand{\doublet}[2]{\left(\begin{array}{c}#1\\#2\end{array}\right)}
\newcommand{\twobytwo}[4]{\left(\begin{array}{cc} #1&#2\\#3&#4\end{array}\right)}

\def\a{\alpha}\def\b{\beta}
\def\g{\gamma}

\def\s{\sigma}

\def\d{\partial}
\def\dag{\dagger}

\def\bV{ {\bf{V}} }\def\bVp{ {\bf{V}}^+ }
\def\bVm{ {\bf{V}}^- }

\newcommand{\Cset}{{\,\,{{{^{_{\pmb{\mid}}}}\kern-.45em{\mathrm C}}}}}
\newcommand{\hF}{\hat F}

\newcommand{\cN}{\mathcal N}

\def\hE{\hat{E}}

\def\Gd{ G^{\dagger} }

\def\ts{ \tilde \sigma }
\def\Gd{ G^{\dagger} }

\numberwithin{equation}{section}

\begin{document}

\allowdisplaybreaks

\renewcommand{\thefootnote}{$\star$}

\renewcommand{\PaperNumber}{058}

\FirstPageHeading

\ShortArticleName{Bifundamental Fuzzy 2-Sphere and Fuzzy Killing Spinors}

\ArticleName{Bifundamental Fuzzy 2-Sphere\\ and Fuzzy Killing Spinors\footnote{This paper is a
contribution to the Special Issue ``Noncommutative Spaces and Fields''. The
full collection is available at
\href{http://www.emis.de/journals/SIGMA/noncommutative.html}{http://www.emis.de/journals/SIGMA/noncommutative.html}}}

\Author{Horatiu NASTASE~$^\dag$ and Constantinos PAPAGEORGAKIS~$^\ddag$}

\AuthorNameForHeading{H. Nastase and C. Papageorgakis}

\Address{$^\dag$~Instituto de F\'{i}sica Te\'{o}rica, UNESP-Universidade Estadual Paulista,\\
\hphantom{$^\dag$}~R. Dr. Bento T. Ferraz 271, Bl. II, Sao Paulo 01140-070, SP, Brazil} 
\EmailD{\href{mailto:nastase@ift.unesp.br}{nastase@ift.unesp.br}}

\Address{$^\ddag$~Department of Mathematics, King's College London,\\
 \hphantom{$^\ddag$}~The Strand, London WC2R 2LS, UK}
\EmailD{\href{mailto:costis.papageorgakis@kcl.ac.uk}{costis.papageorgakis@kcl.ac.uk}}

\ArticleDates{Received March 26, 2010, in f\/inal form July 09, 2010;  Published online July 20, 2010}

\Abstract{We review our construction of a bifundamental version of the fuzzy 2-sphere and its relation to fuzzy Killing spinors,
f\/irst obtained in the context of
the ABJM membrane model. This is shown to be completely equivalent to the usual (adjoint) fuzzy sphere. We discuss the mathematical
details of the bifundamental fuzzy sphere and its f\/ield theory expansion in a model-independent way. We also examine how this new
formulation af\/fects the twisting of the f\/ields, when comparing the f\/ield theory on the fuzzy sphere background with the
compactif\/ication of the `deconstructed' (higher dimensional) f\/ield theory.}

\Keywords{noncommutative geometry; fuzzy sphere; f\/ield theory}

\Classification{81T75; 81T30}

\renewcommand{\thefootnote}{\arabic{footnote}}
\setcounter{footnote}{0}

\section{Introduction and Motivation}

Noncommutative geometry is a tool that f\/inds numerous applications in the description of a~wide range of physical systems. A celebrated example appearing in String Theory is in terms of the polarisation phenomenon discovered by Myers, in which  $N$ D$p$-branes in the presence of transverse Ramond--Ramond f\/lux  distribute themselves onto the surface of a higher-dimensional sphere~\cite{Myers:1999ps}. The physics of the simplest case  are captured by a $\U(N)$ theory, with the  solution involving fuzzy 2-spheres \cite{Hoppe:1982,Hoppe:1988gk,Madore:1991bw}. These are related to families of Hermitian matrices obeying the $\SU(2)$ algebra
 \begin{gather}\label{SU(2)condition}
[ X^{i} , X^{j} ] = 2i \epsilon^{ijk } X^k .
\end{gather}
The $X^i$ enter the physics as ground state solutions to the equations of motion via (\ref{SU(2)condition}). Then their commutator action on the
space of all $N \times N $ matrices organises the matrices into representations of $\SU(2) \simeq \SO(3)$.  An important aspect of the geometry
of the fuzzy 2-sphere involves the construction of fuzzy (matrix) spherical harmonics in $\SU(2)$ representations, which approach the space of
all classical $S^2$ spherical harmonics in the limit of large matrices \cite{Hoppe:1988gk}. This construction of fuzzy spherical harmonics allows
the analysis of f\/luctuations in a non-Abelian theory of D$p$-branes to be expressed at large $N$ in terms of an Abelian higher dimensional theory.
This describes a~D$(p+2)$ brane  wrapping the sphere, with $N$ units of worldvolume magnetic f\/lux. At f\/inite $N$ the higher dimensional theory
becomes a noncommutative $\U(1)$ with a UV cutof\/f \cite{Iso:2001mg,Dasgupta:2002hx,Dasgupta:2002ru,Papageorgakis:2005xr}.

In this article we review a novel realisation of the fuzzy 2-sphere involving bifundamental matrices. The objects that crucially enter the construction are discrete versions of Killing spinors on the sphere \cite{Nastase:2009ny,Nastase:2009zu}\footnote{The work in   \cite{Nastase:2009ny} was carried out in collaboration with S.~Ramgoolam.}. The motivation is similar to the above and comes from the study of the model recently discovered by Aharony, Bergman, Jaf\/feris and Maldacena (ABJM) describing the dynamics of multiple parallel M2-branes on a $\mathbb Z_k$ M-theory orbifold \cite{Aharony:2008ug}, which followed the initial investigations of Bagger--Lambert and Gustavsson (BLG) \cite{Bagger:2006sk,Bagger:2007jr,Bagger:2007vi,Gustavsson:2007vu}.

The ABJM theory is an $\cN = 6$ superconformal Chern--Simons-matter theory with $\SO(6)$ R-symmetry and gauge group $\U(N) \times \U(\bar N) $.  The two Chern--Simons (CS) terms have equal but opposite levels $(k,-k)$ and the matter f\/ields transform in the bifundamental representation.  One can use the inverse CS level $1/k$ as a coupling constant to perform perturbative calculations. At $k=1$ the theory is strongly coupled and describes membranes in f\/lat space. For $k=1,2$ the supersymmetry and R-symmetry are nonperturbatively enhanced to $\cN = 8$ and $\SO(8)$ respectively \cite{Aharony:2008ug,Gustavsson:2009pm,Kwon:2009ar}. It is then possible to use this action to investigate aspects of the $\mathrm{AdS}_4/\mathrm{CFT}_3$ duality, with the role of the 't Hooft coupling played by $\lambda = \frac{N}{k}$. The action of the $\mathbb Z_k$ orbifold on the $\mathbb C^4$ space transverse to the M2's is such that taking $k\to\infty$ corresponds to shrinking the radius of the M-theory circle and entering a IIA string theory regime.

Of particular interest are the ground-state solutions of the maximally supersymmetric massive deformation of ABJM found by Gomis, Rodr\'iguez-G\'omez,
Van Raamsdonk and Verlinde (GRVV) \cite{Gomis:2008vc}\footnote{The mass-deformed theory was also presented in \cite{Hosomichi:2008jb}.}. The theory
still has a $\U(N) \times \U(\bar N) $ gauge group and  $\mathcal N=6$ supersymmetry but conformal invariance is lost and the R-symmetry is broken down
to $\SU(2)\times \SU(2)\times \U(1)$. Its vacua are expected to describe a conf\/iguration of M2-branes blowing up into spherical M5-branes in the
presence of transverse f\/lux through a generalisation of the Myers ef\/fect. At $k=1$ these solutions should have a dual description in terms of the
$\sfrac{1}{2}$-BPS M-theory geometries with f\/lux found in \cite{Bena:2004jw,Lin:2004nb}.

Interestingly, the matrix part of the above ground-state equation is given by the following simple relation, which we will refer to as the GRVV
algebra\footnote{The same def\/ining matrix equation appears while looking for BPS funnel   solutions in the undeformed ABJM theory and f\/irst appeared
as such in \cite{Terashima:2008sy}. Its relation to the   M2--M5 system was also investigated in \cite{Hanaki:2008cu}.}:
\begin{gather}
 G^\a=G^\a G^\dagger_\b G^\b-G^\b G^\dagger_\b G^\a,
\label{equatio}
\end{gather}
where $ G^\alpha$ are $N\times \bar N$ and $ G_\alpha^\dagger$ are $\bar N \times N$ matrices respectively. Given that the Myers ef\/fect for
the M2-M5 system should employ a 3-dimensional surface, one might initially expect  this to represent the def\/ining relation for a fuzzy 3-sphere.
Moreover, the explicit irreducible solutions of (\ref{equatio}) satisfy $G^\a G^\dagger_\a=1$, which seems to suggest the desired fuzzy 3-sphere structure.

However, we will see that the requisite $\SO(4)$ R-symmetry, that would be needed for the existing fuzzy $S^3$ construction of Guralnik and Ramgoolam (GR) \cite{Guralnik:2000pb,Ramgoolam:2001zx,Ramgoolam:2002wb}, is absent in this case. As was also shown in \cite{Nastase:2009ny}, the GR fuzzy $S^3$ construction implies the following algebra
\begin{gather}
\epsilon^{mnpq}X^+_nX^-_pX^+_q = 2\left(\frac{(r+1)(r+3)+1}{r+2}\right)X^+_m,\nonumber\\
\epsilon^{mnpq}X^-_nX^+_pX^-_q = 2\left(\frac{(r+1)(r+3)+1}{r+2}\right)X^-_m\label{defrel} ,
 \end{gather}
which must be supplemented with the sphere condition
\begin{gather*}
 X_mX_m=X_m^+X_m^-+X_m^-X_m^+=\frac{(r+1)(r+3)}{2}\equiv N 
 \end{gather*}
 and the constraints
\begin{gather*}
 X^+_mX^+_n=X^-_mX^-_n=0. 
 \end{gather*}
Here $r$ def\/ines a representation of $\SO(4)\simeq \SU(2)\times \SU(2)$ by ${\cal R}_r^+$ and ${\cal R}_r^-$, with labels $(\frac{r+1}{2}, \frac{r-1}{2})$ and $(\frac{r-1}{2}, \frac{r+1}{2})$ respectively for the two groups, and  the $X_m^\pm$ are constructed from gamma mat\-rices. Even
though the algebra (\ref{defrel}) looks similar to the GRVV  algebra~(\ref{equatio}), they coincide only in the `fuzziest' case with $r=1$,
i.e.\ the BLG ${\cal A}_4$-algebra, which in the Van Raamsdonk $\SU(2)\times \SU(2)$ reformulation \cite{VanRaamsdonk:2008ft} is
\begin{gather*}
R^2X^m=-ik\epsilon^{mnpq}X^nX^{\dagger p}X^q .
\end{gather*}
This fact suggests that equation~(\ref{equatio}) does not describe a fuzzy $S^3$. Furthermore, the perturbative calculations that lead to the above equation are valid at large $k$, where the ABJM theory is describing IIA String Theory instead of M-theory and as a result a D2--D4 bound state in some nontrivial background.

In the following, we will review how solutions to equation~(\ref{equatio}) actually correspond to a fuzzy 2-sphere, albeit in a realisation involving bifundamental instead of the usual adjoint matrices, by constructing the full spectrum of spherical harmonics. This is equivalent to the usual construction in terms of the $\SU(2)$ algebra (\ref{SU(2)condition}). In fact there is a one-to-one correspondence between the representations of the $\SU(2)$ algebra $X_i$ and the representations in terms of bifundamental matrices. We will also show how the matrices $ G^\alpha$, which are solutions of the GRVV algebra up to gauge transformations, correspond to fuzzy Killing spinors on the sphere, recovering the usual Killing spinors in the large $N$ limit.

The purpose of this article is to present the mathematical aspects of the above construction in a completely model-independent way and highlight some of its features simply starting from~(\ref{equatio}). The reader who is interested in the full background and calculations in the context of the ABJM model is referred to \cite{Nastase:2009ny,Nastase:2009zu}, where an analysis of small f\/luctuations around the ground-states at large $N$, $k$ showed that they can be organised in terms of a $\U(1)$ theory on $\mathbb R^{2,1}\times S^2$, consistent with an interpretation as a D4-brane in Type IIA. The full 3-sphere expected from M-theory then appeared as the large $N$, $k=1$ limit of a fuzzy Hopf f\/ibration, $S^1/\mathbb Z_k\hookrightarrow S_F^3/\mathbb Z_k\stackrel{\pi}{\rightarrow} S_F^2$, in which the M-theory circle $S^1/\mathbb Z_k$ is f\/ibred over the noncommutative sphere base, $S^2_F$.

We also discuss how this bifundamental formulation af\/fects the twisting of the f\/ields
when `deconstructing' a higher dimensional f\/ield theory. This is achieved by studying the f\/ield theory around a fuzzy sphere background, where the twisting is necessary in order to preserve supersymmetry. Even though the twisting is usually described in the context of compactifying the higher dimensional `deconstructed' theory,  we show how this naturally  arises from the bifundamental fuzzy sphere f\/ield theory point of view.

The rest of this paper is organised as follows. In Section~\ref{constructing} we give the
harmonic decomposition of the GRVV matrices and relate them to the fuzzy supersphere. In Section~\ref{equivalence} we present a one-to-one map between the adjoint and bifundamental fuzzy sphere constructions, while in Section~\ref{Hopffibration} we establish that connection in terms of the fuzzy Hopf f\/ibration and def\/ine the fuzzy version of Killing spinors on $S^2$. We then discuss  the resulting `deconstruction' of higher dimensional
f\/ield theories on the 2-sphere, specif\/ically the issue of twisting of the f\/ields in order to preserve supersymmetry.
In Section~\ref{deconstruction} we review the process and discuss the dif\/ferences between the adjoint and bifundamental cases, while in Section~\ref{supersymmetric} we brief\/ly discuss a particular application by summarising the results of \cite{Nastase:2009ny,Nastase:2009zu}. We conclude with some closing remarks in Section~\ref{conclusions}.

\section{Constructing the f\/luctuation expansion}\label{constructing}

{\bf Notation.} In this section, we will denote by $k$, $l$, $m$, $n$ the matrix indices/indices of states in a~vector space, while keeping $i,j=1,\dots,3$ as vector indices on the fuzzy $S^2$. We will also use~$j$ for the $\SU(2)$ spin and $Y_{lm}$ for $S^2$ spherical harmonics, following the standard notation. The distinction should be clear by the context.

\subsection{Ground-state matrices and symmetries}

We begin by writing the ground-state solutions to (\ref{equatio}), found in \cite{Gomis:2008vc} and given by
\begin{gather}
\big( G^1\big)_{m,n }    = \sqrt { m- 1 }  \delta_{m,n},\nonumber\\
\big( G^2\big)_{m,n} = \sqrt { ( N-m ) }  \delta_{ m+1 , n },\nonumber\\
\big(G_1^{\dagger} \big)_{m,n} = \sqrt { m-1}  \delta_{m,n},\nonumber\\
\big( G_2^{\dagger}\big)_{m,n} = \sqrt { (N-n ) }  \delta_{ n+1 , m } .\label{BPSmatrices}
\end{gather}
Using the decomposition of the above complex  into real coordinates
\begin{gather}
G^1=X^1+iX^2   , \qquad G^2=X^3+iX^4,
\label{complex}
\end{gather}
one easily sees that  these satisfy
\begin{gather*}
\sum_{p=1}^4 X_pX^p\equiv G^\a G_\a^\dagger=N-1 ,
\end{gather*}
which at f\/irst glance would seem to indicate a fuzzy $S^3$ structure. However, note that in the above $G^1=G_1^\dagger$ for the ground-state solution. With the help of (\ref{complex}) this results in $X_2=0$, which is instead indicative of a fuzzy $S^2$.

As usual in the case of fuzzy sphere constructions, the matrices $G^\a$ will be used to construct both the symmetry operators (as bilinears in $G$, $G^\dagger$, also acting on $G^\a$ themselves) as well as fuzzy coordinates, used to expand in terms of spherical harmonics on the fuzzy sphere.

\subsubsection[$GG^\dagger$ relations]{$\boldsymbol{GG^\dagger}$ relations}\label{2.1.1}

As a f\/irst step towards uncovering the $S^2$ structure we  calculate the $G \Gd$ bilinears
\begin{gather*}
  \big( G^1 G_1^{\dagger} \big)_{m,n} = (m-1)  \delta_{mn},\nonumber\\
  \big( G^2 \Gd_2 \big)_{mn} = ( N-m )  \delta_{mn},\nonumber\\
  \big( G^{1} \Gd_2 \big)_{ mn} = \sqrt { ( m-1) ( N-m+1) }  \delta_{ m , n+1 },\nonumber\\
  \big( G^2 \Gd_1 \big)_{mn} = \sqrt { ( N-m ) m }  \delta_{m+1 , n },\nonumber\\
   \big( G^\alpha G_\alpha^{\dagger} \big)_{mn} = ( N-1 )   \delta_{mn} .
\end{gather*}
Def\/ining $ J^{\a}_{ \b } = G^{\a } \Gd_{ \b }  $ we get the following commutation relation
\begin{gather*}
[ J^{\a}_{\b}   , J^{ \mu }_{ \nu }  ] = \delta^{\mu }_{\b } J^{\a }_{\nu}
     - \delta^{\a}_{\nu} J^{\mu }_{\b } .
\end{gather*}
These are commutation relations of the generators of $\U(2) $.
Then the $ J_{i} = ( \tilde \s_i )^{\a}_{\b} J_{\a}^{\b} $
 are the generators of $\SU(2)$ that result in the usual formulation
of the fuzzy\footnote{Note that more correctly, we should have written
${J^\a}_\b=G^\a G_\b^\dagger$ and
\[
J_i={(\tilde\s_i)^\a}_\b{J^\b}_\a={(\s_i)_\b}^\a {J^\b}_\a ,
\]
but in the following we will stick to the notation $J^\a_\b$. The kind of matrix multiplication  that one has will be made clear from the context.} $S^2$,  in terms of the algebra
\begin{gather}\label{normsu2}
 [ J_i , J_j ] = 2i \epsilon_{ijk} J_k  .
\end{gather}
The trace $J\equiv J^\a_\a=N-1$ is a trivial $\U(1)\simeq\U(2)/\SU(2)$ generator, commuting with everything else.

\subsubsection[$G^\dagger G$  relations]{$\boldsymbol{G^\dagger G}$  relations}\label{2.1.2}

Next, we calculate the $ \Gd G$  combinations
\begin{gather*}
  \big(  \Gd_1  G^1  \big)_{mn}   = (m-1) \delta_{ mn},\nonumber\\
  \big(  \Gd_2  G^2  \big)_{mn}  = (N-m+1) \delta_{mn} - N \delta_{m1} \delta_{n1},\nonumber\\
 \big( \Gd_1  G^2  \big)_{mn} = \sqrt { ( m-1 ) ( N-m ) } \delta_{m+1 , n },\nonumber\\
 \big( \Gd_2 G^1 \big)_{mn} = \sqrt { ( m-2)( N-m+1 ) } \delta_{ m,n+1 },\nonumber\\
 \big( \Gd_\alpha G^\alpha \big)_{mn}  = N \delta_{mn } - N \delta_{ m1 } \delta_{n1 }
\end{gather*}
and def\/ine $ \bar J_{\a}^{\b} = \Gd_{\a} G^{\b}$. The commutation relations for the above then form another copy of~$\U(2)$
\begin{gather*}
 [ \bar J^{\a}_{\b} , \bar J^{\mu }_{\nu } ] = \delta_{\b}^{\mu } \bar
 J^{\a}_{\nu }  - \delta_{\nu  }^{\a} \bar J^{\mu }_{\b }
\end{gather*}
and similarly,
$ \bar{J}_{i} = ( \tilde \s_i )^{\a}_{\b} \bar{J}_{\a}^{\b} $ once again satisfy the usual $\SU(2)$ algebra\footnote{Again, note that we should have written ${\bar{J}_\a\,}^\b=\Gd_\a G^\b$ which emphasises that for $\bar{J}$, the
lower index is the f\/irst matrix index, and
\[
\bar{J}_i={(\tilde{\s}_i)^\a}_\b {\bar{J}_\a\,}^\b={(\sigma_i)_\b}^\a{\bar{J}_\a\,}^\b ,
\]
which emphasises that as matrices, the $\bar{J}_i$ are def\/ined with the Pauli matrices, whereas $J_i$ was def\/ined with
their transpose. However, we will again keep the notation $\bar{J}^\b_\a$.}, for another fuzzy $S^2$
\[
 [ \bar{J}_i , \bar{J}_j ] = 2i \epsilon_{ijk} \bar{J}_k .
\]

The trace
\begin{gather}
(\bar{J})_{mn}=(\bar{J}^\a_\a)_{mn}=N\delta_{mn}-N\delta_{m1}\delta_{n1},\label{barJ}
\end{gather}
which is a $\U(1)\simeq\U(2)/\SU(2)$ generator, commutes with the $\SU(2)$ generators $\bar{J}_i$, though as a~matrix does not commute with the generators $J^1_2$ and $J^2_1$ of the f\/irst set of $\SU(2)$ generators.

At this point, it seems that we have two $\SU(2)$'s, i.e.\ $\SO(4)\simeq\SU(2)\times \SU(2)$ as expected for a 3-sphere, even though we have not yet shown that these are proper space symmetries: We have only found that the $J, \bar J $ satisfy a certain symmetry algebra. In fact, we will next see that these are not independent but rather combine into a single $\SU(2)$.

\subsubsection[Symmetry acting on bifundamental $(N,\bar{N})$ matrices]{Symmetry acting on bifundamental $\boldsymbol{(N,\bar{N})}$ matrices}

All the $(N,\bar{N})$ bifundamental  scalar matrices are of the type
$ G$, $G G^{\dagger} G$, $G G^{\dagger} G G^{\dagger} G$, $\ldots  $. The simplest such terms are the $G^\a$ matrices themselves, the action of the symmetry generators on which we will next investigate.

It is easy to check that the matrices $G^\a$ satisfy
\begin{gather*}
G^1 G^{\dagger}_2 G^2 - G^2 G^{\dagger}_2 G^1  =  G^1 ,\qquad
 G^2 G^{\dagger}_1 G^1 - G^1 G^{\dagger}_1 G^2  =  G^2 .
\end{gather*}
Using the def\/initions of $J_i$ and $\bar J_i$, we f\/ind
\begin{gather}
J_i G^{\a } - G^{\a}  \bar J_i =  ( \tilde \sigma_i )^{\a}_{\b} G^{\b} . \label{offdiagtrans}
\end{gather}
The $G^1$, $G^2 $
transform like the $ (1,0) $ and $ (0,1)$ column
vectors of the spin-$\frac{1}{2}$ representation with
the $J$'s and $ \bar J $'s matrices in the
$ \mathfrak u(N) \times\mathfrak u( \bar N)$ Lie algebra.

By taking Hermitian conjugates in (\ref{offdiagtrans}), we f\/ind that the antibifundamental f\/ields,
$G^\dagger_\a$, transform as
\begin{gather}\label{offdiagtransdagger}
G^{\dagger}_{\a}  J_i - \bar J_i G^{\dagger}_{\a}  =
 G^{\dagger}_{\b}   ( \tilde \sigma_i )^{\b}_{\a } .
\end{gather}

Therefore the $G^{\a}$, $G^\dagger_{\a}$ form a representation when acted by both $J_i $ and $\bar J_i $, but neither symmetry by itself gives a representation for $G^\a$, $G^\dagger_\alpha$.  This means that the geometry we will be constructing from bifundamental f\/luctuation modes has a single $\SU(2)$ symmetry, as opposed to two.  Equations~(\ref{offdiagtrans})  and (\ref{offdiagtransdagger}) imply relations giving transformations between $J_i$ and $\bar J_i$, thus showing they represent the same symmetry
\[
 \Gd_{\g} J_i  G^{\g} =   ( N+1 ) \bar J_i ,\qquad
 G^{\g} \bar J_i  \Gd_{\g }  =  ( N-2) J_i  .
\]

Writing the action of the full $\SU(2) \times \U(1) $ on the $G^\a$, including the $\U(1)$ trace $\bar{J}$, we obtain
\begin{gather}
 J^{\a}_{\b} G^{\g } - G^{\g} \bar J^{\a}_{\b}
 = \delta^{\g}_{\b} G^{ \a} - \delta^{\a}_{\b} G^{\g}, \label{u2}
\end{gather}
while taking Hermitian conjugates of (\ref{u2}) we obtain the $\U(2)$ transformation of $\Gd_\a$,
\[
\bar J^{\a}_{\b} \Gd_{\g} - \Gd_{\g}  J^{\a}_{\b}
=  - \delta^{\a}_{\g} \Gd_{ \b} + \delta^{\a}_{\b} \Gd_{\g} .
\]
The consequence of the above equations is that $G^{ \a } $ has charge
$1$ under the $\U(1)$ generator $ \bar J $. Thus
a global $\U(1)$ symmetry  action on $G^{\a}$ does not leave
the solution invariant, and we need to combine with the action of
$ \bar J$ from the gauge group to obtain an invariance.

We next turn to the construction of fuzzy spherical harmonics out of~$G^\a$.

\subsection[Fuzzy $S^2$ harmonics  from $\U(N)\times\U(\bar N)$ with bifundamentals]{Fuzzy $\boldsymbol{S^2}$ harmonics  from $\boldsymbol{\U(N)\times\U(\bar N)}$ with bifundamentals}\label{decomposition}

All bifundamental matrices of $\U(N)\times\U(\bar N)$, are maps between two dif\/ferent vector spaces. On the other hand, products of the bilinears $G\Gd$ and $G^\dagger G$ are adjoint matrices mapping back to the same vector space. Thus, the basis of `fuzzy spherical harmonics' on our fuzzy sphere will be constructed out of all possible combinations: $\U(N)$ adjoints like $G\Gd$, $G\Gd G\Gd$, $\ldots$, $\U(\bar N)$ adjoints like $\Gd G$, $\Gd G\Gd G$, $\ldots$, and bifundamentals like $G$, $G\Gd G$, $\ldots$ and $\Gd$, $\Gd G \Gd$, $\ldots$.

\subsubsection{The adjoint of $\U(N)$}

Matrices like $ G \Gd $ act on an $N$ dimensional vector space that we call $ {\bf V}^+ $. Thus the space of linear maps from $ {\bf V}^+ $ back
to itself, $ End ( { \bf V }_+ ) $, is the adjoint of the $\U(N)$ factor in the $ \U(N) \times \U( \bar N ) $ gauge group and $G \Gd$ are
examples of matrices belonging to it.  The space $ {\bf V}^+ $ forms an irreducible representation of $ \SU(2) $ of spin $ j =\frac{ N-1}{ 2 }$,
denoted by $V_N$
\begin{gather*}
 \bVp = V_N  .
\end{gather*}
The set of all operators of the form $  G \Gd$,  $G \Gd G \Gd$, $\ldots  $ belong in ${\rm End} ( { \bf V }_+ ) $ and can be expanded in
a basis of `fuzzy spherical harmonics'
def\/ined using the $\SU(2)$ structure. Through the $\SU(2) $ generators $J_i$ we can form the fuzzy spherical harmonics as
\begin{gather*}
  Y^{0}  = 1,\qquad
  Y^1_{ i } = J_i,\qquad
  Y^2_{ ((i_1 i_2 ))} = J_{ {((} i_1 } J_{ i_2 {))} },\qquad
  Y^l_{ ((i_1 \cdots i_l )) } = J_{ (( i_1 } \cdots J_{ i_l  )) } .
\end{gather*}
In the above, the brackets $ (( i_1 \cdots i_l )) $ denote traceless symmetrisation. The complete space of
$N\times N $  matrices can be expanded in the fuzzy
spherical harmonics with $ 0 \le l \le 2j = N-1  $.  One indeed checks that
\[
N^2 = \sum_{l=0}^{2j} ( 2l+1 ) .
\]
Then, a general matrix in the adjoint of $\U(N)$ can be expanded as
\[
A=\sum_{l=0}^{N-1}\sum_{m=-l}^l a^{lm} Y_{lm}(J_i) ,
\]
where
\[
 Y_{lm}(J_i) = \sum_i f_{lm}^{((i_1\cdots i_l))} J_{i_1}\cdots J_{i_l} .
\]
The $ Y_{lm}(J_i)$ become the usual spherical harmonics in the `classical' limit, when $N\to\infty$ and the cut-of\/f in the angular momentum is removed.

In conclusion, all the matrices of $\U(N)$ can be organised into irreps of $\SU(2)$ constructed out of $J_i$, which form the fuzzy spherical harmonics $ Y_{lm}(J_i)$.

\subsubsection[The adjoint of $\U ( \bar  N ) $]{The adjoint of $\boldsymbol{\U ( \bar  N )}$}

In a fashion similar to the $\U(N)$ case, the matrices $ \Gd G$, $\Gd G \Gd G$, $\ldots $, are linear endomorphisms of $ {\bf V}^-$.  These matrices are in the adjoint of the $ \U ( \bar N ) $ factor of the $ \U(N) \times \U ( \bar N ) $ gauge group, and will be organised into irreps of the $\SU(2)$ constructed out of $\bar{J}_i$.

However,  we now have a new operator: We have already noticed in (\ref{barJ}) that the $\U(1)$ generator $\bar{J}$ is nontrivial. We can express it as
\[
\bar J = \Gd_{\a} G^{\a} = N - N \bar E_{11} .
\]
This means that $ {\rm End} ( {\bf V}_- )$ contains in addition to the identity matrix, the matrix $ \bar E_{11} $ which is invariant under $ \SU(2) $. If we label the basis states in $ { \bf V }^- $ as $ | e^-_{k} \rangle  $
 with $ k =1,\dots , N $, then $ \bar E_{11} = |e^{-}_1\rangle \langle e^{-}_1 | $. This in turn means that ${\bf V}^- $ is a reducible representation
\begin{gather*}
 {\bf V}^- = V^-_{N-1 } \oplus  V^-_{1 } .
\end{gather*}
The f\/irst direct summand is the irrep of $\SU(2) $ with dimension $N-1$ while the second is a one-dimensional irrep.
Indeed, one checks that the $ \bar J_i $'s annihilate the state $ | e^-_1 \rangle $, which is necessary for the
identif\/ication with the one-dimensional irrep to make sense.

As a result, the space $ {\rm End} ( {\bf V}^- )$ decomposes as follows
\[
 {\rm End} ( {\bf V}^- ) = {\rm End} ( V^-_{N-1 } ) \oplus {\rm End} ( V^-_1 )
\oplus {\rm Hom} ( V^-_{N-1 } , V^-_1 ) \oplus {\rm Hom} ( V^-_{1} , V^-_{N-1} ) ,
\]
that is, the matrices split as $M_{\mu\nu}=(M_{ij},M_{11},M_{1i},M_{i1})$.
The f\/irst summand has a decomposition in terms of another set of fuzzy spherical harmonics
\[
 Y_{lm}(\bar J_i) =  \sum_i  f_{lm}^{((i_1\cdots i_l))} \bar J_{i_1}\cdots \bar J_{i_l} ,
\]
 for $ l $ going from $0$ to $N-2$, since
\[
(N-1)^2 = \sum_{l=0}^{N-2}  ( 2l + 1 )  .
\]
This gives only  matrices in the $(N-1)$ block, i.e.\ the  ${\rm End} ( V^-_{N-1 } )$.
The second summand is just one matrix transforming in the trivial irrep, $\bar E_{11}$.
The remaining two $N-1$ dimensional spaces of matrices cannot be expressed
as products of $ \bar J_i$. They are spanned by
\[
\bar E_{1k}  =  | e^-_1 \rangle \langle e^-_k |  \equiv   g^{--}_{1k }, \qquad
\bar E_{k1}  =  | e^-_k \rangle \langle e^-_1 |   \equiv    g^{--}_{k 1} ,
\]
which are like spherical harmonics for ${\rm Hom} ( V^-_{N-1 } , V^-_1 ) \oplus {\rm Hom} ( V^-_{1} , V^-_{N-1} )$.
They transform in the $N-1$ dimensional irrep
of $\SU(2) $ under the adjoint action of $ \bar J_i $ and are zero mode eigenfunctions of the $\U(1)$ symmetry operator $\bar J$.

Therefore, one can expand a general matrix in the adjoint of $\U ( \bar N ) $ as
\begin{gather}
\bar{A} = \bar a_{0} \bar E_{11} + \sum_{ l=0}^{N-2}\sum_{m=-l}^l \bar a_{l  m }
Y_{lm}(\bar J_i) + \sum_{k=2 }^N b_{k } g^{--}_{1k } +
\sum_{k=2}^N \bar b_k g^{--}_{k 1} \label{ubarnexp}
\end{gather}
and note that we could have replaced $\bar{E}_{11}$ with the $\U(1)$ generator
$\bar{J}$ by redef\/ining $\bar{a}_0$ and $\bar{a}_{00}$.

{\sloppy
In the large $N $ limit the $ Y_{lm}(\bar J_i)$ become the ordinary spherical harmonics of $S^2$, just like~$Y_{lm}(J_i)$.  There are order $N^2$ of these modes, which is appropriate as the fuzzy $S^2$ can roughly be thought of as a 2-dimensional space with each dimension discretised in $N$ units.  The mode~$\bar a_0$,~$b_k$ and~$\bar b_k$ can be neglected at large $N$, as they have much less than~$N^2$ degrees of freedom.

}

\subsubsection[$\SU(2)$ harmonic decomposition of bifundamental matrices]{$\boldsymbol{\SU(2)}$ harmonic decomposition of bifundamental matrices}\label{harmdec}

As in the case of the $\U(\bar{N})$ matrices, the bifundamental matrices of the form $ G$, $G \Gd G$, $\ldots $ giving physical f\/luctuating f\/ields, are not enough to completely f\/ill ${\rm Hom} ( \bV^- , \bV^+ ) $.  Given the decomposition $ \bVm = V_{N-1}^- \oplus V_{1 }^- $, we decompose ${\rm Hom} ( \bV^- , \bV^+ ) $ as
\[
{\rm Hom} ( \bV^-  , \bV^+ ) = {\rm Hom} ( V_{N-1}^- , V^+_N  ) \oplus {\rm Hom} ( V_1^- ,V^+_N  ) ,
\]
i.e.~the matrices $M_{\mu\nu}$ as $(M_{i\nu},M_{1\nu})$.
The f\/irst summand has dimension $ N ( N-1) $, while
the  second has dimension $ N $ and forms an irreducible representation of $\SU(2)$.

Since the $ V_{N-1}^- $ and  $V^+_N$ are irreps of $\SU(2)$
we can label the states with the eigenvalue of~$\bar J_3$,~$J_3  $
respectively. Given our normalisation of the $\SU(2)$ generators
in~(\ref{normsu2}), the usual spin is $ \frac{ J_3^{\max} }{ 2 } $.
The matrices in ${\rm Hom} ( V_{N-1}^- , V^+_N )$ are of the form $ | e^+_m \rangle\langle e^-_n | $, where $ m = \frac{ -N+1}{ 2 } ,\frac{ -N +3}{2 } ,\dots ,
\frac{N-1}{2}  $,  $ n = \frac{-N +2}{2} , \frac{-N+4}{2} ,\dots , \frac{N-2}{2}  $  denote the eigenvalues of $ \frac{J_3}{2}$.
  These are spanned by matrices of the form $ G ( \bar J_{i_1} ) ( \bar J_{i_2} ) \cdots ( \bar J_{i_l} )$, i.e.\ the matrix $G$ times matrices in ${\rm End}(V_{N-1}^-)$.

The operators in ${\rm Hom} ( V_{N-1}^- , V^+_N )$ transform
in representations of spin  $ l+\frac{1}{2}$ for $ l = 0,\dots , N-2 $.
The dimensions of these representations correctly add up to
\[
\sum_{ l=0 }^{ N-2} ( 2l + 2  )  = N ( N - 1 ) .
\]
This then gives the $\SU(2) $ decomposition of  ${\rm Hom} ( V_{N-1}^- , V^+_N  )$ as
\[
{\rm Hom} ( V_{N-1}^- , V^+_N  ) = \bigoplus_{l=0}^{N-2}  V_{l+ 1/2 } .
\]

On the other hand, matrices $ |e^+_{k}\rangle \langle e^-_1 | \equiv \hE_{k 1 }\in {\rm Hom} ( V_1^- , V^+_N ) $ cannot be written in terms
of the $G$'s and $G^\dagger$'s alone, because $G^{\a} $ acting on $| e^{-}_1\rangle $ gives zero. The index $k$ runs over the $N$ states
in $ \bV^+$.  Here $\hE_{k1}$ are eigenfunctions of the operator $\bar { E}_{11} $ with unit charge,
\[
\hE_{k1}\bar {E}_{11}=\hE_{k1} .
\]

Combining all of the above, the bifundamental f\/luctuations $ r^{ \alpha } $
 can be expanded as follows
\[
r^\a=  r^{ \a }_{\b} G^\b  +\sum_{k=1}^Nt^\a_k\hE_{k1} ,
\]
with
\[
  r^{ \a }_{\b} = \sum_{l=0}^{N-2}\sum_{m=-l}^l(r^{lm})^\a_\b  Y_{lm}(J_i) .
\]
We further decompose $ r^\a_\b $ into a trace and a traceless part
 and def\/ine
\begin{gather*}
s^\a_\b  =   r^{ \a }_{\b} - \frac{1}{2 } \delta^{\a}_{\b } r^{\g}_{\g},\qquad
r =    r^{\g}_{\g},\qquad
T^{ \a }  =  t^{ \alpha}_{ k } \hE_{k 1} .
\end{gather*}
Thus the complete expansion of $r^\a$ is given simply in terms of
\begin{gather}
r^{ \a } = r G^{ \a } + s^{\a}_{ \b  } G^{ \b }+ T^{ \alpha}. \label{fluctuati}
\end{gather}
We could equivalently have written
\begin{gather*}
r^\a=\sum_{l=0}^{N-2}\sum_{m=-l}^l(r^{lm})^\a_\b G^\b  Y_{lm}(\bar J_i)+\sum_{k=1}^Nt^\a_k\hE_{k1}
\end{gather*}
using the spherical harmonics in $ \bar J$ in   (\ref{ubarnexp}).
In the following, we will choose, without loss of generality,  to work with (\ref{fluctuati}).

Until now we have focused on matrices in ${\rm Hom} ( \bV^- , \bV^+ ) $ but the case of ${\rm Hom} ( \bV^+ , \bV^- ) $ is similar. The matrices $ \Gd$, $\Gd G \Gd$, $\ldots  $ will also form a representation of  $\SU(2)$ given by $\bar{J}\sim \Gd G $, times a $\Gd$ matrix. Once again one needs to add an extra $ T^{\dagger}_{ \a } = ( t^{ \alpha}_{ k } )^* \hF_{1 k} $ f\/luctuation in order to express the matrices $\hF_{ 1 k } \equiv |e^-_{1}\rangle \langle e^+_k | \in {\rm Hom}(V_N^+,V_1^-)$. In fact, the result for the complete f\/luctuating f\/ield can be obtained by taking a Hermitian conjugate of (\ref{fluctuati}), yielding
\begin{gather*}
r^{ \dagger }_{ \alpha } =  \Gd_{ \a } r + \Gd_{ \b } s^{\b}_{\a} + T^{\dagger}_{\alpha}. 
\end{gather*}

\subsection{Fuzzy superalgebra}

The matrices $G^\a$ and $J_i$ can be neatly  packaged into supermatrices which form a representation of the orthosymplectic Lie
superalgebra $\text{OSp}(1|2)$. The supermatrix is nothing but the embedding  of the $N\times \bar N$
matrices into $\U(2N)$. The adjoint f\/ields live in the  `even subspace', while the bifundamentals in the `odd subspace'.  For a generic supermatrix
\[
M = \twobytwo{A}{B}{C}{D}
\]
the superadjoint operation is
\[
M^\ddagger = \twobytwo{A^\dagger}{C^\dagger}{-B^\dagger}{D^\dagger} .
\]
For Hermitian supermatrices this is
\[
X = \twobytwo{A}{B}{-B^\dagger}{D} ,
\]
with $A = A^\dagger$ and $D = D^\dagger$ \cite{Hasebe:2004yp}. This gives the def\/inition of the supermatrices
\[
{\bf J}_i = \twobytwo{J_i}{0}{0}{\bar J_i}\qquad\textrm{and}\qquad {\bf J}_\alpha = \twobytwo{0}{\sqrt N  G_\alpha}{-\sqrt N G^\dagger_\alpha}{0} ,
\]
where we raise and lower indices as $G_\alpha = \epsilon_{\alpha\beta} G^\beta$,
with $\epsilon = i \ts_2 = -i\s_2$. Then the SU(2) algebra together with the relation (\ref{offdiagtrans})
and the def\/inition of $J_i$, $\bar J_i$ result in the following (anti)commutation relations
\begin{gather*}
[{\bf J}_i,{\bf J}_j]  =  2 i \epsilon_{ijk} {\bf J}_k, \qquad
[{\bf J}_i,{\bf J}_\alpha]  =  {(\ts_i)_\alpha}_\beta {\bf J}^\beta,\nonumber\\
\{{\bf J_\alpha},{\bf J_\beta}\}  =  - (\tilde \sigma_i)_{\alpha\beta} {\bf J}_i= - (i\tilde \sigma_2 \tilde \sigma_i)_{\alpha\beta} {\bf J}_i,
\end{gather*}
which is the def\/ining superalgebra $\text{OSp}(1|2)$ for the fuzzy {\it supersphere} of \cite{Grosse}.

It is known that the only irreducible representations of $\text{OSp}(1|2)$ split into the spin-$j$ plus the spin-$(j-\frac{1}{2})$
representations of $\SU(2)$, which correspond {\em precisely} to the irreducible representation for the $J_i$ (spin $j$) and $\bar J_i$
(spin $j-1/2$) that we are considering
here\footnote{See for instance Appendix C of \cite{Hasebe:2004yp}. The general spin-$j$ is the $J_i$ representation constructed from the GRVV matrices, while the general spin $j-\frac{1}{2}$ is the $\bar J_i$ representation constructed from the GRVV matrices.}.

As a result, the most general representations of the fuzzy superalgebra, including $G^\a$ besides~$J_i$,~$\bar J_i$, coincide with the most general representations of the two copies of $\SU(2)$.  This points to the fact that perhaps the representations in terms of $G^\a$ are equivalent to the representations of $\SU(2)$. Next we will see that this is indeed the case.

\section{Equivalence of fuzzy sphere constructions}\label{equivalence}

We now prove that our new def\/inition of the fuzzy 2-sphere in terms of bifundamentals is equivalent to the usual
def\/inition in terms of adjoint representations of the $\SU(2)$ algebra.

The ABJM bifundamental scalars are interpreted as Matrix Theory ($N\times N$) versions of Euclidean coordinates. Accordingly,
for our fuzzy space solution in the large $N$-limit one writes $G^\a\rightarrow \sqrt{N}g^\a$, with $g^\a$
some commuting classical objects, to be identif\/ied and better understood in due course.
In that limit, and similarly writing $J_i \rightarrow N x_i$, $\bar J_i \rightarrow N \bar x_i$, one has from Sections~\ref{2.1.1} and~\ref{2.1.2}
that the coordinates
\begin{gather}
x_i = {(\ts_i)^\a}_\b g^\b g^*_\a,\qquad
\bar x_i =  {(\ts_i)^\a}_\b g^*_\a g^\b\label{hopf}
\end{gather}
are two versions of the same Euclidean coordinate on the 2-sphere, $x_i\simeq \bar x_i$.

In the above construction the 2-sphere coordinates $x_i, \bar x_i$  are invariant under multiplication of the classical objects $g^\a$ by a
$\U(1)$ phase, thus we can def\/ine objects $\tilde{g}^\a$ \emph{modulo} such a phase, i.e.\ $g^\a=e^{i\a(\vec{x})}\tilde g^\a$.  The GRVV matrices
(\ref{BPSmatrices}), that from now on we will denote by
$\tilde{G}^\a$ instead of~$G^\a$, are fuzzy versions of representatives of $\tilde g^\a$, chosen such that $\tilde{g}^1=\tilde{g}_1^\dagger$
(one could of course have chosen a dif\/ferent representative for $\tilde{g}^\a$ such that $\tilde{g}^2=\tilde{g}_2^\dagger$ instead).

In terms of the $ g^\alpha$, equation~(\ref{hopf}) is the usual Hopf map from the 3-sphere $ g^\a g^\dagger_\a=1$ onto the 2-sphere $x_ix_i=1$,
as we will further discuss in the next section. In this picture, the phase is simply the coordinate on the $\U(1)$ f\/ibre of the Hopf f\/ibration,
while the $\tilde g^\alpha$'s are coordinates on the $S^2$ base. While $g^\a$ are complex coordinates acted upon by $\SU(2)$, the $\tilde g^\a$
are real objects acted upon by the spinor representation of $\SO(2)$, so they can be thought of as Lorentz spinors in two dimensions, i.e.\
spinors on the 2-sphere.

The fuzzy version of the full Hopf map, $J_i={(\ts_i)^\a}_\b G^\b G^\dagger_\a$, can be given either using $G^\a= U \tilde G^\a $ or
$G^\alpha = \tilde{\hat   G}^\alpha\hat{U}$. The $U$ and $\hat{U}$ are unitary matrices that can themselves be expanded in terms of
fuzzy spherical harmonics
\begin{gather*}
 U=\sum_{lm} U_{lm} Y_{lm}(J_i) ,
 \end{gather*} with $U U^\dagger = \hat U \hat U^\dagger =1$, implying that in the
large-$N$ limit $(U,\hat U)\rightarrow e^{i\a(\vec{x})}$.

That means that by extracting a unitary matrix from the left or the right of $G^\a$, i.e.\ modulo a unitary matrix, the resulting algebra for $\tilde{G}^\a$
\begin{gather}
-\tilde{G}^\a=\tilde{G}^\b \tilde{G}^\dagger_\b \tilde{G}^\a-\tilde{G}^\a \tilde{G}^\dagger_\b \tilde{G}^\b
\label{algebra}
\end{gather}
should then be exactly equivalent to the usual $\SU(2)$ algebra that appears in the adjoint construction: Both should give the same description of the fuzzy 2-sphere. We would next like to prove this equivalence for all possible representations.

\subsection{Representations}

We f\/irst note that the irreducible representations of the algebra (\ref{algebra}), given by the matri\-ces~(\ref{BPSmatrices}), indeed give the most general irreducible representations of SU(2). Def\/ining $J_{\pm}=J_1\pm iJ_2$, $\bar J_{\pm}=\bar J_1\pm i \bar J_2$, we obtain from~(\ref{BPSmatrices}) that
\begin{gather*}
(J_+)_{m,m-1} = 2\sqrt{(m-1)(N-m+1)}=2\alpha_{\frac{N-1}{2},m-\frac{N+1}{2}},\\
(J_-)_{n-1,n} = 2\sqrt{(n-1)(N-n+1)}=2\alpha_{\frac{N-1}{2},n-\frac{N+1}{2}},\\
(J_3)_{mn} = 2\left(m-\frac{N+1}{2}\right)\delta_{mn}
\end{gather*}
and
\begin{gather*}
(\bar J_+)_{m,m-1} = 2\sqrt{(m-2)(N-m+1)}=2\alpha_{\frac{N-2}{2},m-\frac{N+2}{2}},\\
(\bar J_-)_{n-1,n} = 2\sqrt{(n-2)(N-n+1)}=2\alpha_{\frac{N-2}{2},n-\frac{N+2}{2}},\\
(\bar J_3)_{mn} = 2\left(m-\frac{N+2}{2}\right)\delta_{mn}+N\delta_{m1}\delta_{n1} ,
\end{gather*}
whereas the general spin-$j$ representation of $\SU(2)$ is
\begin{gather*}
 (J_+)_{m,m-1}=\alpha_{j,m},\qquad
 (J_-)_{n-1,n}=\alpha_{j,n},\qquad
 (J_3)_{mn}=m\delta_{mn}
\end{gather*}
(and the rest zero), where
\begin{gather*}
\alpha_{jm}\equiv \sqrt{(j+m)(j-m+1)}
\end{gather*}
and $m\in -j,\dots ,+j$ takes $2j+1$ values. Thus the representation for $J_i$ is indeed the most general
$N=2j+1$ dimensional representation, and since $(\bar J_+)_{11}=(\bar J_-)_{11}=(\bar J_3)_{11}=0$, the representation
for $\bar J_i$ is also the most general $(N-1)=2(j-\frac{1}{2})+1$ dimensional representation.

We still have the $\U(1)$ generators completing the  $\U(2)$ symmetry,
which in the case of the irreducible GRVV matrices $\tilde G^\a$
are diagonal and give the fuzzy sphere constraint $\tilde G^\a\tilde G^\dagger_\a\propto \one$, $\tilde G^\dagger _\a \tilde G^\a\propto \one$,
\begin{gather*}
 J={J^1}_1+{J^2}_2=(N-1)\delta_{mn},\qquad
 \bar J ={\bar J_1\,}^1+{\bar J_2\,}^2=N\delta_{mn}-N\delta_{m1}\delta_{n1} ,
\end{gather*}
where again $(\bar J)_{11}=0$, since $\bar J_i$ is in the $N-1\times N-1$ dimensional representation:
The element $E_{11}=\delta_{m1}\delta_{n1}$ is a special operator, so the f\/irst element of the vector space on which
it acts is also special, i.e.\ ${\bf V}^-=V^-_{N-1}\oplus V^-_1$.

Moving to reducible representations of $\SU(2)$, the Casimir operator $\vec{J}^2=J_iJ_i$ giving the fuzzy sphere constraint is diagonal,
with blocks proportional to the identity. The analogous object that  gives the fuzzy sphere constraint in our construction is the
operator $J=G^\a G^\dagger_\a$. Indeed, in the case of reducible matrices modulo unitary transformations, $\tilde{G}^\a$,
we f\/ind (in the same way as for $\vec{J}^2=J_i J_i$ for the $\SU(2)$ algebra)
\begin{gather}
J=\text{diag}((N_1-1)\one_{N_1 \times N_1},(N_2-1)\one_{N_2 \times N_2},\dots )\label{j}
\end{gather}
and similarly for $\bar J=G^\dagger_\a G^\a$
\begin{gather}
\bar J=\text{diag}\big(N_1\big(1-E^{(1)}_{11}\big)\one_{N_1 \times N_1}, N_2\big(1-E^{(2)}_{11}\big)\one_{N_2 \times N_2},\dots \big).
\label{jbar}
\end{gather}

\subsection[GRVV algebra $\rightarrow \SU(2)$ algebra]{GRVV algebra $\boldsymbol{\rightarrow \SU(2)}$ algebra}

For this direction of the implementation one does not need to consider the particular representations of the algebra; the matrices $\tilde{G}^\a$ will be kept as arbitrary solutions. We def\/ine as before, but now for an arbitrary solution $G^\a$,
\begin{gather}
G^\a G^\dagger_\b\equiv {J^\a}_\b\equiv\frac{J_i {(\ts_i)^\a}_\b +J\delta^\a_\b }{2}.\label{definiti}
\end{gather}
Using the GRVV algebra it is straightforward to verify that $G^\a G^\dagger_\a\equiv J$ commutes with $J_k$.

Multiplying (\ref{algebra}) from the right by ${(\ts_k)^\gamma}_\a G^\dagger_\gamma$, one obtains
\[
-J_k = G^\b G^\dagger _\b J_k-{J^\a}_\b {J^\b}_\gamma {(\ts_k)^\gamma}_\a .
\]
Using the def\/inition  for the ${J^\a}_\b$ factors in (\ref{definiti}) and the relation $[J,J_k]=0$, one
arrives at
\[
-J_k=\frac{i}{2}\epsilon_{ijk} J_i J_j ,
\]
which is just the usual SU(2) algebra.

It is also possible to def\/ine
\[
 G^\dagger_\a G^\b\equiv {\bar J_\a\,}^\b\equiv \frac{\bar   J_i{(\ts_i)^\b}_\a+\bar J\delta^\b_\a}{2}
\]
 and similarly obtain  $[\bar J,\bar J_k]=0$. By multiplying (\ref{algebra}) from the left by ${(\ts_k)^\gamma}_\a G^\dagger_\gamma$, we get in a~similar way
\[
 -\bar J_k=\frac{i}{2}\epsilon_{ijk}\bar J_i \bar J_j .
\]

Thus the general SU(2) algebras for $J_i$ and $\bar J_i$ indeed follow immediately from (\ref{algebra}) without restricting to the irreducible GRVV matrices.

\subsection[$\SU(2)$ algebra $\rightarrow$ GRVV algebra]{$\boldsymbol{\SU(2)}$ algebra $\boldsymbol{\rightarrow}$ GRVV algebra}

This direction of the implementation is {\it a priori} more problematic since, as we have already seen, the representations of $J_i$ and $\bar J_i$ are not independent. For the irreducible case in particular, $V_N^+$ is replaced by the representation $V^-_{N-1}\oplus V^-_1$, so we need to generalise this identif\/ication to reducible representations in order to prove our result. As we will obtain this relation at the end of this section and it should have been the starting point of the proof, we will close with some comments summarising the complete logic.

We will f\/irst try to understand
the classical limit. The Hopf f\/ibration (\ref{hopf}) can be rewritten, together with the normalisation
condition, as
\[
 g^\alpha g^*_\beta = \frac{1}{2} \big[x_i {(\ts_i)^\alpha}_\beta + \delta^\alpha_\beta\big]   .
\]

By extracting a phase out of $g^\a$, we should obtain the variables $\tilde{g}^\a$ on $S^2$ instead of $S^3$.
Indeed, the above equations can be solved  for $g^\a$ by
\begin{gather}
g^\alpha = \doublet{g^1}{g^2} = \frac{e^{i \phi}}{\sqrt{2(1+x_3)}}{\doublet{1+x_3}{x_1-ix_2}}=e^{i\phi}\tilde{g}^\a,
\label{tildeg}
\end{gather}
where $e^{i\phi}$ is an arbitrary phase.

In the fuzzy case $G^\a$ and $G^\dagger_\b$ do not commute, and there are two dif\/ferent kinds of equations corresponding to $J_i$ and $\bar J_i$,
\begin{gather}
G^\alpha G^\dagger_\beta  \equiv  \frac{1}{2} \big[J_i {(\ts_i)^\alpha}_\beta + \delta^\alpha_\beta J\big],\qquad
G^\dagger_\beta  G^\alpha  \equiv  \frac{1}{2} \big[\bar J_i {(\ts_i)^\alpha}_\beta + \delta^\alpha_\beta \bar J\big] .
\label{reverse}
\end{gather}
We also impose  that $[J,J_k]=0$, $[\bar J,\bar J_k]=0$, so that $J$ and $\bar J$ are diagonal and proportional to the identity in the irreducible components of $J_i$.

We solve the f\/irst set of equations in (\ref{reverse}) by  writing $G^1G^\dagger_1=\frac{1}{2}(J+J_3)$, for which the
most general solution is $G_1 = T U$, with $T$  a Hermitian and $U$ a unitary matrix.
Since $J+J_3$ is real and diagonal, by def\/ining
\begin{gather*}
T = \frac{1}{\sqrt 2} (J+ J_3 )^{1/2}
\end{gather*}
we obtain
\begin{gather}\label{320}
G^\alpha = \doublet{G^1}{G^2} = {\doublet{J+J_3}{J_1-iJ_2}}\frac{T^{-1}}{2}U_{N\times N}
=\tilde{G}^\alpha U_{N\times N} .
\end{gather}
Thus $\tilde{G}^\alpha$ is also completely determined by $J_i$, $J$.

Similarly,  the second set of equations in (\ref{reverse}) can be solved by  considering $G^\dagger_1 G^1=\frac{1}{2} (\bar J+\bar J_3)$, for which the most general solution is $G^1=\hat{U}\tilde{T}$, where as before
\[
\tilde{T} = \frac{1}{\sqrt 2}\Big(\bar J+ \bar J_3\Big)^{1/2} ,
\]
to obtain
\begin{gather}\label{322}
G^\alpha = \doublet{G^1}{G^2} = \hat{U}_{\bar N\times\bar N}
\frac{\tilde{T}^{-1}}{2}{\doublet{\bar J+\bar J_3}{\bar J_1-i\bar J_2}}=\hat{U}\tilde{\hat G}^\a .
\end{gather}
Thus $\tilde{\hat G}^\a$ is completely determined by $\bar J_i$, $\bar J$.

Comparing the two formulae for $G^\a$ we see that they are compatible if and only if
\begin{gather}
\hat{U}=TU\tilde{T}^{-1}\qquad\text{and}\qquad
\bar J_1-i\bar J_2=\tilde{T}^2U^{-1}T^{-1}(J_1-iJ_2)T^{-1}U ,\label{equiva}
\end{gather}
where $U$ is an arbitrary unitary matrix. These equations def\/ine an identif\/ication between the two representations of $\SU(2)$, in terms of $J_i$ and $\bar J_i$, needed  in order to establish the equivalence with the GRVV matrices.

We now analyse the equivalence for specif\/ic representations.
For the irreducible representations of $\SU(2)$, we def\/ine $\bar J_i$ from $J_i$ as before ($V_N^+\rightarrow V_{N-1}^-\oplus V_1^-$) and $J=(N-1)\one_{N \times N}$, $\bar J=N(1-E_{11})\one_{N   \times N}$.  For reducible representations of $\SU(2)$, $J_i$ can be split such that $J_3$ is block-diagonal, with various irreps added on the diagonal.  One must then take $J$ and $\bar J$ of the form in (\ref{j}) and (\ref{jbar}). The condition (\ref{equiva}) is solved by $U=1$ and $J_1$, $J_2$ block diagonal, with the blocks being the irreps of dimensions $N_1,N_2,N_3,\dots $, and the $\bar J_1$, $\bar J_2$ being also block diagonal, but where each $N_k\times N_k$ irrep block is replaced with the $(N_k-1)\times (N_k-1)$ irrep block, plus an $E^{(k)}_{11}$, just as for the GRVV matrices.

We can hence  summarise the proof {\it a posteriori} in the following steps:
\begin{enumerate}\itemsep=0pt
\item Start  with $J_i$ ($i = 1,2,3$) in the reducible representation of $\SU(2)$, i.e.\ block diagonal with the blocks being irreps of dimensions $N_1, N_2, N_3, \dots $.
\item Take $J = G^\alpha G^\dagger_\alpha$ and $\bar J = G^\dagger_\alpha G^\alpha$ as in (\ref{j}) and (\ref{jbar}) since these are necessary conditions for the $G^\alpha$ to satisfy the GRVV algebra. The condition $[J, J_k]=0$ is used here.

\item The $\bar J_i$ are  completely determined (up to conventions) from $J_i$, $J$ and $\bar J$ by (\ref{equiva}) and the condition $[\bar J, \bar J_k]=0$.

\item The $\tilde G^\alpha$ are then uniquely determined by (\ref{320}), while the $\tilde{\hat G}^\alpha$ by (\ref{322}).

\item The $\tilde G^\alpha$ and $\tilde{\hat G}^\alpha$ def\/ined as above indeed satisfy the GRVV algebra.

\end{enumerate}

\section{Fuzzy Hopf f\/ibration and fuzzy Killing spinors}\label{Hopffibration}

Having established the equivalence between the adjoint (usual) and the bifundamental (in terms of $\tilde{G}^\a$) formulations of the fuzzy $S^2$ we turn towards ascribing an interpretation to the mat\-ri\-ces~$\tilde{G}^\a$ themselves.

\subsection{Hopf f\/ibration interpretation}

One such interpretation was alluded to already in  (\ref{complex}), where the fuzzy (matrix) coordinates $G^\a$ were treated as complex spacetime coordinates.  The irreducible GRVV matrices satisfy $ \tilde G^1 \tilde G^{\dagger}_1 + \tilde G^2 \tilde G^{\dagger}_2 = N-1 $ and $\tilde G^1 = \tilde G^{\dagger}_1 $.  The f\/irst relation suggests a fuzzy 3-sphere, but the second is an extra constraint which reduces the geometry to a 2d one. This is in agreement with the fuzzy $S^2$ equivalence that we already established in the previous section. The matrices~$\tilde G^\a$ are viewed as representatives when modding out the $\U(N)$ symmetry, and the condition $\tilde G^1=\tilde G^\dagger_1$ amounts to a choice of representative of the equivalence class.

The construction  of the fuzzy $S^2$ in usual (Euclidean) coordinates
was obtained by
\begin{gather*}
 J_i  =  ( \tilde \sigma_i )^{\a}_{\b} G^{\b} G^{\dagger}_{\a},\\
 x_i  =  \frac{ J_i}{\sqrt{ N^2-1} }\  \Rightarrow
\left\{ \begin{array}{l}
\displaystyle x_1 = \frac{ J_1}{\sqrt{ N^2-1} } = \frac{ 1}{\sqrt{ N^2-1} } \big( G^1 G^{\dagger}_2 +
 G^2 G^{\dagger}_1\big),   \vspace{1mm}\\
 \displaystyle  x_2 = \frac{J_2}{\sqrt{ N^2-1} } =
\frac{i}{\sqrt{ N^2-1} } \big(   G^1 G^{\dagger}_2 -  G^2 G^{\dagger}_1 \big),  \vspace{1mm}\\
\displaystyle x_3 = \frac{J_3}{\sqrt{ N^2-1} } = \frac{1}{\sqrt{ N^2-1}} \big( G^1 G_1^{\dagger} - G^2 G_2^{\dagger} \big)  ,
\end{array} \right.  \\
\frac{G^\a}{\sqrt{N}} \rightarrow  g^\a
\end{gather*}
and we already stated that the relation between $g^\a$ and $x_i$  is the  classical Hopf map  $S^3\stackrel{\pi}{\rightarrow} S^2$,~(\ref{hopf}).

Indeed, the description of the Hopf map in classical geometry
is given as follows: One starts with Cartesian coordinates $X_1$, $X_2$, $X_3$, $X_4 $ on the unit $S^3$ with
\[
 X_1^2 + X_2^2 + X_3^2 + X_4^2 =1
\]
and then  goes to complex variables $Z^1 = X_1 + i X_2$, $Z^2 = X_3 + i X_4 $, satisfying $Z^\alpha Z^*_\alpha = 1$. The Hopf map def\/ines
Cartesian coordinates on the unit $S^2$ base of the f\/ibration by
\begin{gather}\label{classHopf}
x_i = ( \tilde \sigma_i )^{ \a}_{\b} Z^{ \b} Z^*_{\a}  ,
\end{gather}
which is invariant under an $S^1$ f\/ibre def\/ined by multiplication of $Z^\a$ by a phase. The $x_i$ are Euclidean coordinates on an $S^2$ since
\[
 x_i x_i =  ( \tilde \sigma_i )^{ \a}_{\b}
( \tilde \sigma_i )^{ \mu  }_{\nu } Z^{ \b} Z^*_{\a}
  Z^{ \nu } Z^*_{\mu } = 1
\]
and this identif\/ies $Z^\alpha \equiv g^\alpha$ from above.

Let us now work in the opposite direction, starting from the classical limit and discretising the geometry by demoting the Hopf map (\ref{classHopf}) from classical coordinates to f\/inite matrices. We need matrices for $Z^{\a}$ which we call $ G^{\alpha }$.  The coordinates $x_i$ transform in the spin-$1$ representation of $\SU(2)$. If we want to build them from bilinears of the form $G^{ \dagger} G $ we need $G$, $\Gd $ to transform in the spin-$\frac{1}{2}$ representation. We also want a gauge symmetry to extend the $\U(1)$ invariance of $Z^\a$ (the $S^1$ f\/iber of the Hopf map), and for $N$-dimensional matrices $\U(N)$ is the desired complex gauge invariance that plays that role.

In the usual fuzzy 2-sphere, the $x_i$ are operators mapping an irreducible $N$-dimensional $\SU(2)$ representation $ V_N $ to itself. It is possible to do this in an $\SU(2)$-covariant fashion because the tensor product of spin-$1$ with $V_N$ contains $V_N$. Since $G^{\alpha } $ are spin-$\frac{1}{2}$, and $ \frac{ 1}{2 } \otimes V_{N} = V_{N+1} \oplus V_{N-1} $ does not contain $V_N$, we need to work with reducible representations in order to have $G^\a$ map the representation back to itself.  The simplest thing to do would be to consider the representation $ V_N \oplus V_{N-1} $.  The next simplest thing is to work with $ V_{N} \oplus ( V_{N-1} \oplus V_1 ) $ and this possibility is chosen by the GRVV matrices~\cite{Gomis:2008vc} and allows a gauge group $\U(N) \times \U(\bar N)$ which has a $\mathbb Z_2$ symmetry of exchange needed to preserve parity.

{\sloppy
So the  unusual property of the GRVV matrices $\tilde G$, the dif\/ference between $ \bVp = V_N $ and $ \bVm = V_{N-1} \oplus V_1 $ follows from requiring a matrix realisation of the fuzzy $S^2$ base of the Hopf f\/ibration. These in turn lead to the $\SU(2)$ decompositions of $ {\rm End} ( \bVp )$, ${\rm End} ( \bVm )$, ${\rm Hom} ( \bVp , \bVm )$, ${\rm Hom} ( \bVm , \bVp ) $, for the f\/luctuation matrices that we saw in Section~\ref{decomposition}.\footnote{The usual fuzzy $S^2$ has also been discussed in terms of the Hopf f\/ibration,   where the realisation of the $\SU(2)$ generators in terms of bilinears in Heisenberg algebra   oscillators yields an inf\/inite dimensional space which admits various projections to f\/inite $N$   constructions \cite{Balachandran:2005ew}.  In that case the $x_i $ are not bilinears in f\/inite   matrices.}

}

The $x_i$, $G$, $\Gd $ are operators in $ \bVp \oplus \bVm$ which is isomorphic, as a vector space, to ${ \bf V}_N \otimes V_2 $. The endomorphisms of ${ \bf V }_N$ correspond to the fuzzy sphere. The $N $ states of ${ \bf V }_N $ generalise the notion of points on $S^2$ to noncommutative geometry. The 2-dimensional space $V_2$ is invariant under the $\SU(2)$. It is acted on by $G$, $\Gd $ which have charge $ +1$, $-1$ under the $\U(1)$ (corresponding to $(J , \bar J ) $) acting on the f\/ibre of the Hopf f\/ibration, so we also have two points on top of our fuzzy $S^2$.

Since in this subsection we looked at a f\/ibration of $S^3$,
we need to emphasise that the f\/luctuation analysis does not have enough modes to describe the full space of functions on~$S^3$, even if we drop the requirement of $\SO(4)$ covariance and allow for the possibility of an $\SU(2) \times \U(1)$ description.
As we explained above,  the only remnant of the circle in the matrix construction is the multiplicity associated with having states $|+\rangle$, $|-\rangle$ in $ \bVp $ and $ \bVm $. A classical description of the $S^3$ metric as a Hopf f\/ibration contains a coordinate $y$ transverse to the $S^2$. Instead, the matrix f\/luctuations of our solution are mapped to functions on $S^2$ and hence lead to a f\/ield theory on $S^2$.

\subsection{Killing spinor interpretation}\label{killinginterp}

We will close this circle of arguments by interpreting the classical objects $\tilde{g}^\a$, obtained in the large-$N$ limit of $\tilde{G}^\a$, as Killing spinors
and fuzzy Killing spinors on the 2-sphere respectively.

We have seen that in the classical limit the relation between $J_i$ and $G^\a$ becomes the f\/irst Hopf map (\ref{hopf}), and hence can be thought of as its {\it fuzzy} version. However, the above Hopf relation is invariant under multiplication by an arbitrary phase corresponding to shifts on the~$S^1$ f\/ibre, so the objects~$\tilde{g}^\a$ obtained by extracting that phase in~(\ref{tildeg}), i.e.\
\begin{gather}
 \tilde{g}^\a=\frac{1}{\sqrt{2(1+x_3)}}{\doublet{1+x_3}{x_1-ix_2}} ,\label{doubletul}
\end{gather}
are instead def\/ined on the classical $S^2$. In the Hopf f\/ibration, the index of $g^\alpha$ is a spinor index of the global $\SO(3)$ symmetry
for the 2-sphere. By extracting the $S^1$ phase one obtains a real (or rather, subject to a reality condition)
$\tilde{g}^\a$ and the $\a$ can be thought of as describing a (Majorana)
spinor of the $\SO(2)$ local Lorentz invariance on the 2-sphere. We will argue that the latter is related to a Killing spinor. Note that this
type of index identif\/ication easily extends to all even spheres.

In the fuzzy version of (\ref{doubletul}), the $\tilde{G}^\a$ obtained from $G^\alpha$ by extracting a unitary matrix, are real objects
def\/ined on the fuzzy $S^2$. They equal the GRVV matrices in the case of irreducible representations, or
\begin{gather*}
\tilde{G}={\doublet{J+J_3}{J_1-iJ_2}}\frac{T^{-1}}{2}
\end{gather*}
in general.

The standard interpretation, inherited from the examples of the $\SU(2)$ fuzzy 2-sphere and other spaces, is that the matrix indices give rise to the dependence on the sphere coordinates and the index $\a$ is a \emph{global} symmetry index. However,  we have just seen that already in the classical picture one can identify the global symmetry spinor index with the \emph{local} Lorentz spinor index. Therefore we argue that the correct interpretation of the classical limit for $\tilde{G}^\a$ is as a~spinor with both global \emph{and} local Lorentz indices, i.e.\ the Killing spinors on the sphere $\eta^{\a I}$. In the following we will use the index $\a$ interchangeably for the two.

In order to facilitate the comparison with the Killing spinors, we express the classical limit of the $J_i$--$\tilde{G}^\a$ relation as
\begin{gather}
x_i\simeq\bar x_i = {(\sigma_i)_\a}^\b \tilde{g}^\dagger_\b \tilde{g}^\a. \label{classi}
\end{gather}

\subsubsection*{Killing spinors on $\boldsymbol{S^n}$}

We now review some of the key facts about Killing spinors that we will need for our discussion. For more details, we refer the interested reader to e.g.~\cite{vanNieu1983,Eastaugh1985,vanNieuwenhuizen:1984iz,Gunaydin:1984wc,Nastase:1999kf}.

On a general sphere $S^n$, one has Killing spinors satisfying
\begin{gather*}
D_\mu \eta(x)=\pm \frac{i}{2}m\gamma_\mu \eta(x).
\end{gather*}
There are two kinds of
Killing spinors, $\eta^+$ and $\eta^-$, which in even dimensions are related by the chirality matrix, i.e.\ $\gamma_{n+1}$, through $\eta^+=\gamma_{n+1} \eta^-$, as  can be easily checked.
The Killing spinors on~$S^n$ satisfy orthogonality, completeness and a reality condition. The latter depends on the application, sometimes taken to be the \emph{modified} Majorana condition, which mixes (or identif\/ies) the local Lorentz spinor index with the global symmetry spinor index of $S^n$.
For instance, on~$S^4$ the orthogonality and completeness are respectively\footnote{The charge conjugation matrix in $n$ dimensions satisf\/ies in general
\[
C^T=\kappa C,\qquad
\gamma_\mu^T=\lambda C \gamma_\mu C^{-1} ,
\]
where $\kappa=\pm$, $\lambda=\pm$ and it is used to raise/lower indices. The Majorana condition is then given by
\[
\bar \eta =\eta^T C .
\]},
\begin{gather*}
\bar \eta^I \eta^J=\Omega^{IJ} \qquad\text{and}\qquad
\eta^\a_J\bar \eta^J_\b=-\delta_\b^\a ,
\end{gather*}
where the index $I$ is an index in a spinorial representation of the $\SO(n+1)_G$ invariance group of the sphere and the index $\a$ is an index in a spinorial representation of the $\SO(n)_L$ local Lorentz group on the sphere. The indices are then identif\/ied by the \emph{modified} Majorana spinor condition as follows\footnote{For more details   on Majorana spinors and charge conjugation matrices see~\cite{vanNieu1983,VanNieuwenhuizen:1981ae} and the Appendix of \cite{Nastase:1999kf}.}
\[
\bar \eta^I\equiv \big(\eta^I\big)^TC^{(n)}_{-}=-\big(\eta^J\big)^\dagger \gamma_{n+1} \Omega^{IJ} ,
\]
where $\Omega^{IJ} = i \sigma_2 \otimes \one_{\sfrac{n}{2}\times \sfrac{n}{2}}$ is the invariant tensor of $\text{Sp}(\sfrac{n}{2})$, satisfying $\Omega^{IJ}\Omega_{JK} = \delta^I_K$.

The Euclidean coordinates of $S^n$ are bilinear in the Killing spinors
\begin{gather}\label{bilinear}
x_i=(\Gamma_i)_{IJ}\bar \eta^I \gamma_{n+1} \eta^J ,
\end{gather}
where $\eta$ are of a single kind ($+$ or $-$), or equivalently $\bar \eta_+^I\eta_-^J$. In the above the $\Gamma$ are in $\SO(n+1)_G$, while the $\gamma$ in $\SO(n)_L$.

Starting from  Killing spinors on $S^n$, one can construct all the higher spherical harmonics. As seen in equation~(\ref{bilinear}), Euclidean coordinates on the sphere are spinor bilinears. In turn, symmetric traceless products of the $x_i$'s  construct the scalar spherical harmonics $Y^k(x_i)$.\footnote{These are the higher dimensional extensions of the usual spherical harmonics $Y^{lm}(x_i)$ for $S^2$.}  One can also construct the set of spinorial spherical harmonics by acting with an appropriate operator on~$Y^k\eta^I$
\begin{gather*}
\Xi^{k,+} =  [(k+n-1+iD\!\!\!\!/)Y^k]\eta_+,\\
\Xi^{k,-} =  [(k+n-1+iD\!\!\!\!/)Y^k]\eta_-=[(k+1+iD\!\!\!\!/)Y^{k+1}]\eta_+ .
\end{gather*}
Note that in the above the derivatives act only on the scalar harmonics~$Y^k$.

Any spinor on the sphere can be expanded in terms of spinorial spherical harmonics, $\Psi=\sum_k \psi_k\Xi^{k,\pm}$. Consistency imposes that the $\Xi^{k,\pm}$ can only be commuting spinors. The Killing spinors are then themselves {\em commuting} spinors, as they are used to construct the spinorial spherical harmonics.

For higher harmonics the construction extends in a similar way but the formulae are more complicated and, as we will not need them for our discussion, we will not present them here. The interested reader can consult e.g.~\cite{Kim:1985ez}.

\subsubsection*{Killing spinors on $\boldsymbol{S^2}$ and relation between spinors}

For the particular case of the $S^2$, $\gamma_i=\Gamma_i = \sigma_i$ for both the $\SO(2)_L$ and the $\SO(3)_G$ Clif\/ford algebras. Then the two $C$-matrices can be chosen to be: $C_+=-\sigma_1$, giving $\kappa=\lambda=+$, and $C_-=i\sigma_2=\epsilon$, giving $\kappa=\lambda=-$. Note that with these conventions one has $C_-\gamma_3=i\sigma_2\sigma_3=-\sigma_1=C_+$. In the following we will choose the Majorana condition to be def\/ined with respect to $C_-$.

 Equation~(\ref{bilinear}) then gives for $n=2$
\begin{gather}
\bar \eta^I =(\eta^T)^I C_- \ \Rightarrow \ x_i=(\sigma_i)_{IJ}(\eta^T)^I C_- \gamma_3 \eta^J. \label{classiK}
\end{gather}
The orthonormality and completeness conditions for the Killing spinors on $S^2$ are
\begin{gather*}
\bar \eta^I \eta^J=\epsilon^{IJ}\qquad\text{and}\qquad\eta^\a_J \bar \eta^J_\b =-\delta^\a_\b ,
\end{gather*}
while the modif\/ied Majorana condition is
\begin{gather*}
(\eta^J)^\dagger=\epsilon_{IJ}\bar \eta^I\equiv \epsilon_{IJ}(\eta^I)^T C_- .
\end{gather*}
Since $C_-=\epsilon$, by making both indices explicit and by renaming the index $I$ as $\dot\a$ for later use, one also
has
\begin{gather}\label{modifiedeta}
(\eta^{\a\dot\a})^\dagger=\eta_{\a\dot\a}\equiv \epsilon_{\a\b}\epsilon_{\dot\a\dot\b}\eta^{\b\dot\b} .
\end{gather}

Finally, the spinorial spherical harmonics on $S^2$ are
\begin{gather*}
\Xi^{\pm}_{lm}=[(l+1+iD\!\!\!\!/\;)Y_{lm}]\eta_{\pm}
\end{gather*}
and thus the spherical harmonic expansion of an $S^2$-fermion is (writing explicitly the sphere fermionic index $\a$)
\[
\psi^\a=\sum_{lm,\pm} \psi_{lm,\pm}\Xi^{\pm,\a}_{lm}=\sum_{lm,\pm} {[\psi_{lm,\pm}(l+1+iD\!\!\!\!/\;)Y_{lm}]^\a
}_\b\eta_{\pm}^\b .
\]

To construct explicitly the Killing spinor, we must f\/irst def\/ine a matrix $S$, that can be used to relate between the two dif\/ferent kinds
of spinors on~$S^2$, spherical and Euclidean.

On the 2-sphere, one def\/ines the Killing vectors $K_i^a$ such that the adjoint action of the $\SU(2)$ generators on the fuzzy sphere f\/ields becomes a derivation in the large-$N$ limit\footnote{Precise expressions for the Killing vectors as well as a set of useful identities can be found in Appendix~A of~\cite{Nastase:2009zu}.}
\begin{gather*}
[J_i,\cdot]\to 2iK_i^a\d_a=2i\epsilon_{ijk}x_j\d_k .
\end{gather*}
One can then explicitly check that $K_i^a$ produces a Lorentz transformation on the gamma matrices\footnote{A Lorentz transformation on the spinors acts as
${\Lambda^\mu}_\nu \gamma^\nu = S\gamma^\mu S^{-1}$, with~$S$ unitary.}
\begin{gather*}
K_i^a{(\ts_i)^\a}_\b=-e^{am}{\big(S \sigma^m S^{-1}\big)_\b\, }^\a\equiv -{\big(S \gamma^a S^{-1}\big)_\b\,}^\a, 
\end{gather*}
where $e^{am}$ is the vielbein on the sphere and  $S$ is a unitary matrix def\/ining the transformation ($|a|=1$)
\begin{gather*}
S=a\begin{pmatrix}
 -\sin{\frac{\theta}{2}} \,e^{i\phi/2}& \displaystyle -i\cos{\frac{\theta}{2}} \,e^{i\phi/2}\vspace{1mm}\\
\displaystyle \cos{\frac{\theta}{2}} \,e^{-i\phi/2}&\displaystyle  -i\sin{\frac{\theta}{2}} \,e^{-i\phi/2}
\end{pmatrix} .
\end{gather*}
Imposing the (symplectic) reality condition on $S$
\begin{gather}\label{symplectic}
\epsilon_{\a\beta}{\big(S^{-1}\big)^\b}_\gamma \epsilon^{\gamma\delta}={\big(S^T\big)_\a}^\delta=
{S^\delta}_\a ,
\end{gather}
we f\/ix $a=\sqrt{i}^*$ and obtain the relations
\begin{gather}
{\big(S \sigma_i S^{-1}\big)_\a\,}^\b = {\big(S\sigma_i S^{-1}\big)^\b}_\a,\qquad
{\big(S\gamma_3 S^{-1}\big)^\a}_\b  = -x_i{(\ts_i)^\a}_\b,\nonumber\\
{\big(S\gamma_a S^{-1}\big)^\a}_\b  = -h_{ab }K_i^b{(\ts_i)^\a}_\b.\label{gamma3}
\end{gather}

If one has real spinors obeying
\begin{gather*}
(\chi_{\a\dot\a})^\dagger=\chi^{\a\dot\a}\equiv \epsilon^{\a\b}\epsilon^{\dot\a\dot\b}\chi_{\b\dot\b},
\end{gather*}
which was identif\/ied in (\ref{modifiedeta}) as the {\em modified} Majorana spinor condition,
it follows from  (\ref{symplectic}) that rotation by the matrix $S$ preserves this relation, i.e.\
\begin{gather}
((\chi_{\dot\a}S)_\a)^\dagger=\big(S^{-1}\chi^{\dot\a}\big)^\a\equiv -\epsilon^{\dot\a\dot\b}\big(S^{-1}\big)^{\a\b}
\chi_{\b\dot\b}=\epsilon^{\dot\a\dot\b}\epsilon^{\a\b}(\chi_{\dot\b}S)_\b . \label{rotreal}
\end{gather}

We can now def\/ine the explicit form of the Killing spinor
\begin{gather*}
\eta^{I\a}={\big(S^{-1}\big)^\a}_\b \eta_0^{I\b}= \frac{1}{\sqrt{2}} {\big(S^{-1}\big)^\a}_\b \epsilon^{\b I}=\frac{1}{\sqrt{2}}
{S^I}_J \epsilon^{\a J},
\end{gather*}
where in the last equality we used the  (symplectic) reality condition (\ref{symplectic}) on~$S$. From~(\ref{rotreal}) it is clear  that the $\eta^{I\a}$  obey the \emph{modified} Majorana condition. It is then possible to use (\ref{gamma3}) to prove that
\begin{gather*}
x_i=(\sigma_i)_{IJ}\bar \eta^I \gamma_3 \eta^J ,
\end{gather*}
hence verifying that the  $\eta^{I\a}$ are indeed Killing spinors. One can also explicitly check that
\begin{gather*}
D_a\big({\big(S^{-1}\big)^\a}_\b \epsilon^{\b I}\big)=+\frac{i}{2}{(\gamma_a)^\alpha}_\beta {\big(S^{-1}\big)^\beta}_\gamma \epsilon^{\gamma I} ,
\end{gather*}
which in turn means that
\begin{gather*}
\frac{1}{\sqrt{2}} {\big(S^{-1}\big)^\a}_\b \epsilon^{\b I}=\eta_+^{\a I} .
\end{gather*}

\subsubsection*{Identif\/ication with Killing spinor}

Using (\ref{modifiedeta}), we rewrite (\ref{classiK}) as
\begin{gather}\label{432}
x_i={(\sigma_i)^I}_J \big(\eta^I\big)^\dagger \gamma_3\eta^J
={(\ts_i)^I}_J\big(\sqrt{2}P_+\eta^I\big)^\dagger\big(\sqrt{2}P_+\eta^J\big) ,
\end{gather}
where $P_\pm = \frac{1}{2}(1\pm\gamma_3)$. Now comparing (\ref{432}) with (\ref{classi}) one is led to the following natural large-$N$ relation, $\tilde{G}^\a\rightarrow \sqrt{2N} P_+\eta^I$, provided the spinor indices $\a$ and $I$ get identif\/ied, i.e.\
\begin{gather*}
\frac{\tilde{G}^\a}{\sqrt{N}}\equiv \tilde{g}^\a\leftrightarrow \tilde{g}^I\equiv \sqrt{2}P_+\eta^I
={(P_+)^\a}_\b {(S^{-1})^\b}_\g \epsilon^{\g I}
={(P_+)^\a}_\b {S^I}_J \epsilon^{\b J}={S^I}_J{(P_-)^J}_K
\epsilon^{\a K}.
\end{gather*}
Thus, the Weyl projection can be thought of as `removing' either $\a$ or $I$, since only one of the two spinor components is non-zero.

In order to further check this proposed identif\/ication at large-$N$ we now calculate
\begin{gather}
\d_a \big(\sqrt{2}P_+\eta^I\big)= -\frac{i}{2} {\big(S \gamma_a S^{-1}\big)^I}_J
\big(\sqrt{2}P_+\eta^J\big)
+\tilde{T}_a \big(\sqrt{2}P_+\eta^I\big),\label{killisp}
\end{gather}
where $\tilde{T}_\theta=0$ and $\tilde{T}_\phi=\frac{i}{2}\cos\theta$ and
\[
(\d_a S )S^{-1}=-\frac{i}{2}S\gamma_aS^{-1}+S T_a S^{-1}
\]
by explicitly evaluation, with $T_\theta=0$ and $T_\phi=-\frac{i}{2}\cos\theta  \gamma_3$.

This needs to be compared with the analogous result given in equation~(4.48) of \cite{Nastase:2009ny} from the classical limit of the adjoint action of $J_i$ on $\tilde{G}^\a$, i.e.\ from $[J_i,\tilde{G}^\a]$,
\begin{gather}
\d_a \tilde{g}^\a=\frac{i}{2}\hat h_{ab}K_i^b{(\ts_i)^a}_\beta \tilde g^\beta
= -\frac{i}{2}{\big(S\gamma_a S^{-1}\big)^\a}_\b \tilde{g}^\b. \label{classg}
\end{gather}
In \cite{Nastase:2009ny} it was also verif\/ied that the above could reproduce the correct answer for $\d_a x_i$, which can be rewritten as
\[
\d_a x_i=-\frac{i}{2}\tilde{g}^\dagger_\a\big[{(\ts_i)^\a}_\b {\big(S\gamma_a S^{-1}\big)^\b}_\g-{\big(S\gamma_a S^{-1}\big)^\a}_\b {(\ts_i)^\b}_\g\big]
\tilde{g}^\g .
\]

Note that even though there is a dif\/ference between (\ref{killisp}) and (\ref{classg}), given by the purely imaginary term $\tilde{T}_a$
that is proportional to the identity, the two answers for $\d_a x_i$ exactly agree, since in that case the extra contribution cancels.
This extra term is a ref\/lection of a double ambiguity:  First, the extra index $\a$ on $\eta^I$ can be acted upon by matrices,
even though it is Weyl-projected, in ef\/fect multiplying the Weyl-projected $\eta^I$ by a complex number; if the complex number is a phase,
it will not change any expressions where the extra index is contracted, thus we have an ambiguity against multiplication by a phase.
Second, $\tilde{g}^\a$ is just a representative of the reduction of $g^\a$ by an arbitrary phase, so it is itself only def\/ined
up to a phase.  The net ef\/fect is that the identif\/ication of the objects in (\ref{killisp}) and (\ref{classg}) is only up to a phase.
Indeed, locally, near $\phi\simeq 0$, one could write
\[
\tilde{g}^\a e^{\frac{i}{2}\phi \cos\theta} \ \leftrightarrow \ \sqrt{2}P_+\eta^I
\]
but it is not possible to get an explicit expression for the phase over the whole sphere.

\subsection{Generalisations}

On a general $S^{2n}$ some elements of the above analysis of fuzzy Killing spinors carry through.
That is because even though it is possible to write for every $S^{2n}$
\[
x_A=\bar\eta^I (\Gamma_A)_{IJ} \gamma_{2n+1}\eta^J ,
\]
where $\eta^I$ are the Killing spinors, one only has possible fuzzy versions of the  quaternionic and octonionic Hopf maps to match it against,
i.e.\ for $2n=4,8$. We will next f\/ind and interpret the latter in terms of Killing spinors on the corresponding spheres.

\subsubsection*{$\boldsymbol{S^4}$}

The second Hopf map, $S^7\stackrel{\pi}{\rightarrow} S^4$, is related to the quaternionic algebra. Expressing the $S^7$ in terms of
complex coordinates $g^\a$, now with $\a=1,\dots ,4$, the sphere constraint becomes $g^\a g^\dagger_\a=1$ ($g^\a g^\dagger_\a=1\Rightarrow x_A x_A=1$;
$A = 1,\dots ,5$). The map in this case is (see for instance
\cite{Wu:1988py})
\[
x_A=g^\b {(\Gamma_A)^\a}_\b g^\dagger_\a,
\]
with ${(\Gamma_A)^\a}_\b$ the $4\times 4$ $\SO(5)$ gamma matrices\footnote{These are constructed as: $\sigma_1$ and $\sigma_3$ where $1$ is replaced by ${\one}_{2\times 2}$ and
$\sigma_2$ where $i$ is replaced by $i\sigma_1, i\sigma_2,i\sigma_3$.}.  Here we have identif\/ied the spinor index $I$ of $\SO(5)$ with the Lorentz spinor index $\a$ of $\SO(4)$.

Initially, the  $g^\a$'s are complex coordinates acted upon by $\SU(4)$, but projecting down to the base of the Hopf f\/ibration we
replace $g^\a$ in the above formula with real $\tilde{g}^\alpha$'s, instead acted upon by the spinorial representation of $\SO(4)$, i.e.\
by spinors on the 4-sphere. This process is analogous to what we saw for the case of the 2-sphere. Once again, it is possible to
identify $\tilde{g}^\a$ with the Killing spinors, this time on $S^4$.

This suggest that one should also be able to write a spinorial version of the fuzzy 4-sphere
for some bifundamental matrices $\tilde{G}^\a$, satisfying
\[
J_A = \tilde{G}^\b {(\Gamma_A)^\a}_\b \tilde{G}^\dagger_\a, \qquad
\bar J_A =  \tilde{G}^\dagger_\a {(\Gamma_A)^\a}_\b \tilde{G}^\b ,
\]
where $J_A$, $\bar J_A$ generate an $\SO(5)$
spinor rotation on $\tilde G^\a$ by
\[
J_A\tilde G^\a- \tilde G^\a \bar J_A={(\Gamma_A)^\a}_\b\tilde G^\b .
\]
This in turn implies that the fuzzy sphere should be described by the same GRVV algebra as for the $S^2$ case
\[
\tilde G^\a=\tilde G^\a\tilde G^\dagger_\b\tilde G^\b -\tilde G^\b\tilde G^\dagger_\b\tilde G^\a
\]
but now with $\tilde G^\a$ being 4 complex matrices that describe a fuzzy 4-sphere, which poses an interesting possibility that we
will however not further investigate here.

\subsubsection*{$\boldsymbol{S^8}$}
The third Hopf map, $S^{15}\stackrel{\pi}{\rightarrow} S^8$, is related to the octonionic algebra. The $S^{15}$ is expressed now
by the real objects $g^T_\a g^\a=1$, $\a=1,\dots ,16$ that can be split into two groups ($1,\dots ,8$ and $9,\dots ,16$). The Hopf map is expressed  by \cite{Bernevig:2003yz} ($g^T_\a g_\a=1\Rightarrow x_Ax_A=1$)
\[
x_A=g^T_\a(\Gamma_A)^{\a\b}g_\b ,
\]
where $(\Gamma_A)^{\a\b}$ are the $\SO(9)$ gamma-matrices\footnote{The gamma-matrices are constructed
similarly to the $S^4$ case as follows: $\Gamma_i=\begin{pmatrix}0&\lambda_i\\-\lambda_i &0\end{pmatrix}$, $\Gamma_8=
\begin{pmatrix} 0&\one_{8\times 8}\\ \one_{8\times 8}&0\end{pmatrix}$, $\Gamma_9=\begin{pmatrix}\one_{8\times 8}&0\\0&-\one_{8\times
8}\end{pmatrix}$, i.e.\ from $\sigma_2$ with $\lambda_i$ replacing $i$, and from $\sigma_1$ and $\sigma_3$ with
$1$ replaced by $\one_{8\times 8}$. The
$\lambda_i$ satisfy $\{\lambda_i,\lambda_i\}=-2\delta_{ij}$ (similarly to the $i\sigma_i$
in the case of $S^4$) and are constructed from the structure constants of the algebra of the octonions~\cite{Bernevig:2003yz}. An explicit
inversion of the Hopf map is given by $g_\a=[(1+x_9)/2]^{1/2}u_\a$ for $\a=1,\dots ,8$ and $g_\a=[2(1+x_9)]^{-1/2}
(x_8-x_i\lambda_i)u_{\a-8}$ for $\a=9,\dots,16$, with $u_\a$ a real 8-component  $\SO(8)$ spinor satisfying $u^\a u_\a=1$
thus parametrising the $S^7$ f\/ibre.}.
Similarly for the case of the $S^4$ above, even though $g^\a$'s are initially 16-dimensional
variables acted by the spinor representation of $\SO(9)$, one can project down to the base of the Hopf f\/ibration
and replace the $g^\a$'s with real 8-dimensional
objects on the 8-sphere $\tilde{g}^\a$.  Then the $\tilde{g}^\a$'s are identif\/ied with the Killing spinors of $S^8$.

This once again suggests that one should be able to write a spinorial version of the fuzzy 8-sphere for some
bifundamental matrices $\tilde{G}^\a$ satisfying
\[
J_A = \tilde G_\a(\Gamma_A)^{\a\b}\tilde G^T_\b,\qquad
\bar J_A = \tilde G^T_\a(\Gamma_A)^{\a\b}\tilde G_\b ,
\]
where $J_A$, $\bar J_A$ generate an $\SO(9)$
spinor rotation on $\tilde G^\a$ by
\[
J_A\tilde G_\a - \tilde G_\a \bar J_A={(\Gamma_A)_\a}^\b\tilde G_\b
\]
and  implies the same GRVV algebra, but with the $\tilde G^\a$'s now being 16 dimensional real matrices that
describe the fuzzy 8-sphere.

\section{Deconstruction vs.\ twisted compactif\/ication}\label{deconstruction}

We now describe certain changes which occur when `deconstructing' a supersymmetric f\/ield theory on
the bifundamental fuzzy $S^2$, in contrast to the usual $S^2$, and comparing with the compactif\/ied higher-dimensional theory.

The term `deconstruction' was f\/irst coined in \cite{ArkaniHamed:2001ca} for a specif\/ic four-dimensional model but  more generally extends to  creating higher dimensional theories through f\/ield theories with matrix degrees of freedom of high rank. In our particular case, the fuzzy $S^2$ background arises as a~solution in a $d$-dimensional f\/ield theory and f\/luctuations around this background `deconstruct' a~$d+2$-dimensional f\/ield theory. We will focus on the case where the $d+2$-dimensional f\/ield theory compactif\/ied on $S^2$ is supersymmetric.

\subsection[Adjoint fuzzy $S^2$]{Adjoint fuzzy $\boldsymbol{S^2}$}

This construction is familiar in the context of D-branes, though any f\/ield theory with a fuzzy~$S^2$ background will also do. For instance, the example we will follow is \cite{Andrews:2006aw}, where an ${\cal N}=1$ supersymmetric massive $\SU(N)$ gauge theory around a fuzzy $S^2$ background solution, coming from the low energy theory on a stack of D3-branes in some nontrivial background, was identif\/ied with the Maldacena--N\'u\~nez theory of IIB 5-branes with twisted compactif\/ication on $S^2$ \cite{Maldacena:2000mw}. This construction was known to give an ${\cal N}=1$ massive theory after dimensional reduction that can be identif\/ied with the starting point, thus the D3-brane theory around the fuzzy sphere deconstructs the 5-brane theory.

The twisting of the 5-brane f\/ields can be understood both in the compactif\/ication as well as the deconstruction pictures. In compactif\/ication, and for the \cite{Andrews:2006aw} model, it is known from \cite{Bershadsky:1995qy} that in order to preserve supersymmetry on D-branes with curved worldvolumes one needs to twist the various D-brane f\/ields.  Specif\/ically, that means embedding the $S^2$ spin connection, taking values in $\SO(2)\simeq\U(1)$, into the R-symmetry. As a result, the maximal supersymmetry one can obtain after compactif\/ication is ${\cal N}=1$ (corresponding to $\U(1)_R$).  On the other hand, in deconstruction, the need for twisting will instead appear by analysing the kinetic operators of the deconstructed f\/ields.

The brane intuition, though useful, is not necessary, and in the following we will understand the twisting as arising generally from requiring supersymmetry of the dimensionally reduced compactif\/ied theory. This will be matched by looking at the kinetic term diagonalisation of the deconstructed theory.

\subsubsection*{Compactif\/ication}

On a 2-sphere, scalar f\/ields are decomposed in the usual spherical harmonics $Y_{lm}(x_i)=Y_{lm}(\theta,\phi)$ and can thus give massless f\/ields after compactif\/ication (specif\/ically, the $l=0$ modes). However, that is no longer true for spinors and gauge f\/ields. In that case, the harmonic decomposition in terms of $Y_{lm}(x_i)$ must be redef\/ined in order to make explicit the Lorentz properties of spinors and vectors  on the 2-sphere, i.e.\ to make them eigenvectors of their corresponding operators.

Spinors on the sphere are eigenvectors of the total angular momentum $J_i^2$. These are of two types: Eigenvectors $\Omega$
of the orbital angular momentum $L_i^2$ (Cartesian spherical spinors) and eigenvectors $\Upsilon$ of the Dirac operator on the sphere
$-i\hat{\nabla}_{S^2}=-i\hat h^{ab} e^{m}_a\sigma_m\nabla_b$  (spherical basis spinors), whose square is
$R^2(-i\hat{\nabla}_{S^2})^2=J_i^2+\frac{1}{4}$. The two are related by a transformation with a sphere-dependent matrix $S$, already described in
Section \ref{killinginterp}. The former are decomposed in the spinorial spherical harmonics
\[
\Omega^{\hat{\a}}_{jlm}=\sum_{\mu=\pm \sfrac{1}{2}}C(l,\sfrac{1}{2},j;m-\mu, \mu,m)Y_{l,m-\mu}(\theta,\phi)\chi_\mu^{\hat{\a}}  ,
\]
where $j=q_{\pm}=l\pm \frac{1}{2}$ and $\hat{\a}=1,2$, as
\[
\psi^{\hat\a}=\sum_{lm}\psi_{lm}^{(+)}\Omega_{l+\frac{1}{2},lm}^{\hat\a}+\psi_{lm}^{(-)} \Omega_{l-\frac{1}{2},lm}^{\hat\a} .
\]
Both have a minimum mass of $\frac{1}{2R}$, since the Dirac operator squares to $J_i^2+\frac{1}{4}=j(j+1)+\frac{1}{4}$. Similarly,
the vector f\/ields do not simply decompose in $Y_{lm}$'s, but rather in the vector spherical harmonics
\begin{gather*}
  \frac{1}{R}{\bf T}_{jm}=\frac{1}{\sqrt{j(j+1)}}\big[\sin\theta \d_\theta Y_{jm} {\bf \hat\phi}
-\csc\theta \d_\phi Y_{jm}{\bf\hat\theta}\big],\\
  \frac{1}{R}{\bf S}_{jm}=\frac{1}{\sqrt{j(j+1)}}\big[\d_\theta Y_{jm} {\bf \hat\theta}
+\d_\phi Y_{jm}{\bf\hat\phi}\big] ,
\end{gather*}
with $j\geq 1$. It is more enlightening to show the decomposition of the f\/ield strength on the 2-sphere
\begin{gather*}
\frac{1}{R}\csc\theta F_{\theta\phi}=R^2\sum_{lm}F_{lm}\frac{1}{\sqrt{l(l+1)}}\Delta_{S^2}Y_{lm} ,
\end{gather*}
with $l=1,2,\dots $. Thus again only massive and no massless modes are obtained after dimensional reduction \cite{Andrews:2006aw}. Note that as we can see,
the expansion in spinorial or vector spherical harmonics corresponds to redef\/ining the expansion in terms of $Y_{lm}$ (rearranging its coef\/f\/icients).

Therefore in the absence of twisting supersymmetry will be lost after dimensional reduction, since all $S^2$-fermions will be massive but some massless $S^2$-scalars will still remain. Twisting, however, allows for the presence of fermionic twisted-scalars (T-scalars), i.e.\ fermions that are scalars of the twisted $\SO(2)_T$ Lorentz invariance group (with charge $T$), which will stay massless. In this way the number of supersymmetries in the dimensionally reduced theory equals the number of fermionic T-scalars.

One chooses the twisted Lorentz invariance of the sphere as $Q_T=Q_{xy}+Q_A$, where $Q_{xy}$ is the charge under the original Lorentz invariance of the sphere $\SO(2)_{xy}$, and $Q_A$ is the charge under the $\U(1)$ subgroup of R-symmetry.  This is necessary because one needs to identify the $\U(1)$ spin connection (`gauge f\/ield of Lorentz invariance') with a corresponding connection in the R-symmetry subgroup, i.e.\ a gauge f\/ield from the transverse manifold.

An example of an action for twisted f\/ields is provided by the
result of \cite{Bershadsky:1995qy}, for a bosonic T-spinor $\Xi$, fermionic T-scalars $\Lambda$ and T-vectors $g_a$
\begin{gather}
\int\! d^d x d^2\sigma\sqrt{h}\!\left[ -\frac{i}{2}\mu \bar \Lambda \gamma^\mu \d_\mu \Lambda\!-\!\frac{i}{2}\mu
\bar g_a \gamma^\mu \d_\mu g^a\!+\!\mu \omega^{ab} \bar G_{ab}\Lambda\!-\!2\d_\mu \Xi^\dagger \d^\mu \Xi
\!-\!8\Xi^\dagger\big({-}i\hat \nabla_{S^2}\big)^2\Xi\right]\! ,\!\!\label{twistDorey}
\end{gather}
where $\mu$ is the mass parameter, $G_{ab}= \d_{a} g_{b}- \d_{b} g_{a}$ is the f\/ield strength of the fermionic T-vector, and as usual $\omega^{ab}=\frac{1}{\sqrt{g}}\epsilon^{ab}$ is the symplectic form on the sphere.  We note that the kinetic terms in the f\/lat directions ($\mu,\nu$) are given by their bosonic or fermionic nature, while the type of kinetic terms in the sphere directions ($a,b$) are dictated by their T-spin and the number of derivatives on it are again dictated by their statistics (bosons have two derivatives, fermions only one).

These f\/ields are decomposed in spherical harmonics corresponding to their T-charge. Then e.g.\ the fermionic T-scalar can have a massless ($l=0$) mode, which after dimensional reduction  will still be a fermion and give ${\cal N}=1$ supersymmetry.

\subsubsection*{Deconstruction}

To have a fuzzy sphere background of the usual type, we need in the worldvolume theory at least 3 scalar modes $\phi_i$ to satisfy $[\phi_i,\phi_j]=2i\epsilon_{ijk}\phi_k$, but usually there are more. Then  the need for e.g.\ bosonic T-spinors is uncovered by diagonalising the kinetic term for all the scalar f\/luctuations around the fuzzy sphere background.  For instance in \cite{Andrews:2006aw}, there are 6 scalar modes forming 3 complex scalars $\Phi_i$, with f\/luctuations $\delta\Phi_i=a_i+ib_i$ and kinetic term
\begin{gather*}
\int d^d x d^2 \sigma \sqrt{h} \delta\Phi_i^\dagger\big[\big(1+J^2\big)\delta_{ij}-i\epsilon_{ijk}J_k\big]\delta \Phi_j .
\end{gather*}
The (complete set of) eigenvectors of this kinetic operator are given by the vector spherical harmonics $J_i Y_{lm}$ and the spinorial
spherical harmonics $\Omega^{\hat{\a}}_{jlm}$. This kinetic operator is then diagona\-lised by def\/ining T-vectors $n_a$ coming from the
vector spherical harmonics and T-spinors $\xi^{\hat{\a}}$ coming from the spinor spherical harmonics.
When completing this program, the deconstructed action is the same as the compactif\/ied one, e.g.\ for \cite{Andrews:2006aw} one again obtains
the twisted action~(\ref{twistDorey}).

At f\/inite $N$, the matrices are expanded in the fuzzy spherical harmonics $Y_{lm}(J_i)$, becoming the $Y_{lm}(x_i)$ of classical $S^2$, but the above diagonalisation corresponds in the classical limit to re-organising the expansion (this includes a nontrivial action on the coef\/f\/icients of the expansion) to form the spinorial, vector,  etc.\ spherical harmonics.

Thus for the adjoint construction all the f\/ields on the classical $S^2$ appear as limits of functions expanded in the scalar fuzzy spherical harmonics, $Y_{lm}(J_i)$, and the various tensor structures of~$S^2$ f\/ields were made manifest by diagonalising the various kinetic operators.

\subsection[Bifundamental fuzzy $S^2$]{Bifundamental fuzzy $\boldsymbol{S^2}$}

The case of the bifundamental fuzzy $S^2$ is richer. One wants to once again compare with the same compactif\/ication picture. However, the particulars of the deconstruction will be dif\/ferent.

\subsubsection*{Deconstruction}

Here we need a fuzzy sphere background of GRVV type, hence  at least 2 complex scalar modes~$R^\a$ in the worldvolume theory giving the fuzzy sphere background in terms of $R^\a=fG^\a$, with~$G^\a$ satisfying~(\ref{equatio}). The f\/luctuation of this f\/ield will be called $r^\a$.

Performing the deconstruction follows a set of steps similar to the adjoint fuzzy $S^2$, namely one wants to expand in the fuzzy spherical harmonics and in the classical limit reorganise the expansion (acting nontrivially on the coef\/f\/icients of the expansion) to construct the spinor, vector,  etc.\ spherical harmonics. However now there are some subtle points that one needs to take into account. We have two kinds of fuzzy spherical harmonics, $Y_{lm}(J_i)$ and $Y_{lm}(\bar J_i)$, both giving the same $Y_{lm}(x_i)$ in the classical limit. Adjoint f\/ields, e.g.\ the gauge f\/ields, will be decomposed in terms of one or the other according to their respective gauge groups. On the other hand for bifundamental f\/ields one must f\/irst `extract' a bifundamental GRVV matrix, $\tilde G^\a$~or~$\tilde G^\dagger_a$, before one is left with adjoints that can be decomposed in the same way. We detailed this procedure for~$r^\a$ in Section~\ref{harmdec}. The expansion in $Y_{lm}(x_i)$ must be then reorganised as in the usual fuzzy~$S^2$ in order to diagonalise the kinetic operator, thus producing the spinor, vector,  etc.\  spherical harmonics.

The most important dif\/ference is that $\tilde G^\a$ has a spinor index on $S^2$; in particular we saw in Section~\ref{killinginterp} that in the classical limit $\tilde g^\a$ is identif\/ied with a Killing spinor. That means that the operation of `extracting' $\tilde G^a$ corresponds to automatically twisting the f\/ields! Let us make this concrete by considering a specif\/ic example.

  In the mass-deformed ABJM theory, one has besides the $R^\a$ f\/ield a doublet of scalar f\/ields~$Q^{\dot\a}$ with f\/luctuation $q^{\dot\a}$, where $\dot\a$ is an $\SU(2)$ index transverse to the sphere. Thus the $q^{\dot\a}$ start of\/f life as scalars. However,  due to their bifundamental nature, one must f\/irst `extract' $\tilde G^\a\rightarrow \sqrt{N}\tilde g^\a$, by writing $q^{\dot\a}=Q_\a^{\dot\a}\tilde G^\a$. In order to diagonalise the kinetic operator, we perform an S-transformation and construct
\begin{gather}
 \Xi^\alpha_{\dot   \alpha}  = i(P_+ S^{-1} Q_{\dot \alpha})^\alpha + \big(P_- S^{-1} Q_{\dot   \alpha}\big)^\alpha,\label{Qredef}
\end{gather}
 after which the kinetic term becomes the twisted action
\begin{gather}
\label{finaltransverse}
 N^2\int d^3 x d^2\sigma \sqrt {\hat h}\left[ \frac{1}{2} \bar \Xi^{\dot   \alpha} (- i2 \mu\hat\nabla_{S^2})^2 \Xi_{\dot \alpha} -\frac{1}{2} \d_\mu \bar \Xi^{\dot   \alpha}\d^\mu \Xi_{\dot \alpha} - 3 \mu^2\bar \Xi^{\dot \alpha} \Xi_{\dot \alpha} \right] .
\end{gather}

More generally, the functions on the sphere are actually sections of the appropriate bundle: Either ordinary functions, sections of the spinor or the line bundle.  Specif\/ically, anything without an $\alpha$ index is a T-scalar, one $\a$ index implies a T-spinor and two $\a$ indices a T-scalar plus a T-vector in a $( {\bf 1}\oplus {\bf 3})$ decomposition. That is, the $\U(1)_T$ invariance is identif\/ied with the $\SO(2)_L\simeq \U(1)_L$ Lorentz invariance of the sphere, described by the index~$\a$.

In addition to this, an interesting new alternative to the above construction also arises. We can choose to keep $\tilde G^\a$ in the spherical harmonic expansion (by considering it as part of the spherical harmonic in the classical limit). The derivative of the spherical harmonic expansion then includes the derivative of $\tilde g^\a$ given in (\ref{classg}) and  one obtains a fuzzy version of the classical derivative operator
\begin{gather*}
q^\dagger_{\dot{\beta}}J_i-\bar J_i q^\dagger_{\dot{\b}} \ \rightarrow \
2i K_i^a \d_a q^\dagger_{\dot{\b}}+q^\dagger_{\dot\b}x_i .
\end{gather*}
This operator acts on all bifundamental f\/ields, including the ABJM  fermions  $\psi^{\dagger\a}$.
In this new kind of expansion, we recover the usual Lorentz covariant kinetic term. For instance for the scalar f\/ields $q^{\dot\a}$ of ABJM
we obtain (after a rescaling of the f\/ields)
\[
\frac{1}{g_{YM}^2}\int d^3 x d^2 \s \sqrt { h }\big[ {-}\d^A q_{\dot\a}^\dagger \d_A q^{\dot\a}\big] ,
\]
where $A=\mu, a$ is a total (worldvolume + fuzzy sphere) index. The price one pays for this simplicity (compared to~(\ref{finaltransverse})) is however  that the classical $N\rightarrow \infty$ limit of the supersymmetry transformation is very
subtle, since a naive application will relate f\/ields with dif\/ferent f\/inite $N$ gauge structures (bifundamentals with adjoints),
naively implying a gauge-dependent supersymmetry parameter.

But at least formally, by keeping $\tilde G^\a$ inside the spherical harmonic expansion, we obtain an un-twisted, fully supersymmetric version of the action on the whole worldvolume plus the fuzzy sphere.

\section[Supersymmetric D4-brane action on fuzzy $S^2$ from ABJM]{Supersymmetric D4-brane action on fuzzy $\boldsymbol{S^2}$ from ABJM}\label{supersymmetric}

As a concrete application of the whole discussion thus far, we present the f\/inal results for the Lagrangian obtained by studying f\/luctuations around the fuzzy $S^2$ ground-state of the mass-deformed ABJM model.

The f\/luctuating f\/ields are the $r^\a$ scalars forming the fuzzy sphere background, transverse scalars $q^{\dot\a}$,
gauge f\/ields $A_\mu$ and  $\hat A_\mu$, fermions $\psi_\a$ and $\chi_{\dot\a}$.
The spherical harmonic expansion on the fuzzy sphere is for each of the above
\begin{gather*}
r^\a = r\tilde{G}^\a+{s^\a}_\b \tilde{G}^\b=\big[(r)_{lm}\delta^\a_\b+({s^\a}_\b)_{lm}\big]Y_{lm}(J_i)\tilde{G}^\b,
\nonumber\\
q^{\dot\a} = Q^{\dot\a}_\a\tilde{G}^\a=(Q^{\dot\a}_\a)_{lm}Y_{lm}(J_i)\tilde{g}^\a,\nonumber\\
\psi_\a = \tilde{\psi}\tilde{G}_\a+{U_\a}^\b \tilde{G}_\b=\big[(\tilde{\psi})_{lm}\delta_\a^\b+({U_\a}^\b)_{lm}\big]
Y_{lm}(J_i)\tilde{G}_\b,\nonumber\\
\chi_{\dot\a} = \chi_{\dot\a \a}\tilde{G}^\a=(\chi_{\dot\a \a})_{lm}Y_{lm}(J_i)\tilde{G}^\a,\nonumber\\
A_\mu = A_\mu^{lm}Y_{lm}(J_i),\qquad
\hat A_\mu = \hat A_\mu^{lm}Y_{lm}(\bar J_i) ,
\end{gather*}
becoming in the classical limit
\begin{gather*}
r^\a = r\tilde{g}^\a+{s^\a}_\b \tilde{g}^\b=\big[(r)_{lm}\delta^\a_\b+({s^\a}_\b)_{lm}\big]Y_{lm}(x_i)\tilde{g}^\b,
 \\
  q^{\dot\a} = Q^{\dot\a}_\a\tilde{g}^\a=(Q^{\dot\a}_\a)_{lm}Y_{lm}(x_i)\tilde{g}^\a, \\
 \psi_\a = \tilde{\psi}\tilde{g}_\a+{U_\a}^\b \tilde{g}_\b=\big[(\tilde{\psi})_{lm}\delta_\a^\b+({U_\a}^\b)_{lm}\big]
Y_{lm}(x_i)\tilde{g}_\b,\\
 \chi_{\dot\a} = \chi_{\dot\a \a}\tilde{g}^\a=(\chi_{\dot\a \a})_{lm}Y_{lm}(x_i)\tilde{g}^\a,\\
 A_\mu = A_\mu^{lm}Y_{lm}(x_i),\qquad
 \hat A_\mu = \hat A_\mu^{lm}Y_{lm}(x_i) .
\end{gather*}
These can be further redef\/ined as
\begin{gather*}
 {s^\a}_\b {(\ts_i)^\b}_\a=K_i^aA_a+x_i\phi,\qquad
 \Upsilon_{\dot\a}^\a=(P_-S^{-1}\chi_{\dot\a})^\a ,
\end{gather*}
with $A_a$ becoming the sphere component of the gauge f\/ield and $\Phi=2r+\phi$ becoming a scalar, while $2r-\phi$ is `eaten' by the gauge f\/ield
in a Higgs mechanism that takes us from nonpropagating CS gauge f\/ield to propagating YM f\/ield in 3d \cite{Mukhi:2008ux}. The f\/inal supersymmetric version of the
action is then
\begin{gather*} 
S_{\rm phys}  =  \frac{1}{g_{\rm YM}^2}\int d^3 x d^2 \s \sqrt { h }
 \Bigg[{-}\frac{1}{4}  F_{AB}  F^{AB}-\frac{1}{2} \partial_A
\Phi \partial^A\Phi - \frac{\mu^2}{2} \Phi^2 -\d^A q_{\dot\a}^\dagger \d_A q^{\dot\a}
+ \frac{\mu}{2}\; \omega^{ab} F_{ab}\Phi \\
\phantom{S_{\rm phys}  =}{}
+\left(\frac{1}{2}\bar \Upsilon^{\dot \alpha} \tilde D_5  \Upsilon_{\dot\alpha} +
\frac{i}{2} \mu \bar \Upsilon^{\dot\alpha} \Upsilon_{\dot{\a}}+{\rm h.c.}\right)
-(\psi S) \tilde D_5 \big(S^{-1}\psi^\dagger\big) +\frac{i}{2}\mu (\psi S) \big(S^{-1}\psi^\dagger\big) \Bigg] .
\end{gather*}

The twisting of the f\/ields that have a $\tilde G^\a$ in their spherical harmonic expansion is done as follows: First, we twist
by expressing $q^{\dot\a}$ as $Q^{\dot\a}_\a$ and $\psi_\a$ as $\tilde{\psi}$, ${U_\a}^\b$. We then redef\/ine the twisted f\/ields in order to
diagonalise their kinetic operator by further writing $Q_\a^{\dot\a}$ according to (\ref{Qredef}) and
\begin{alignat*}{3}
& {U_\a}^\b = \frac{1}{2}U_i{(\ts_i)_\a}^\b ,\qquad &&
{\bar U_\a\,}^\b=\frac{1}{2} U_i {(\ts_i)_\a}^\b, & \\
& U_i = K_i^a g_a+\hat\psi x_i ,\qquad &&
\bar U_i=K_i^a\bar g_a+\bar{\hat\psi} x_i .
\end{alignat*}

The f\/inal twisted action is
\begin{gather*}
S_{\rm phys}  =  \frac{1}{g_{\rm YM}^2}\int d^3 x d^2 \s \sqrt { h }
 \Bigg[-\frac{1}{4}  F_{AB}  F^{AB}-\frac{1}{2} \partial_A
\Phi \partial^A\Phi - \frac{\mu^2}{2} \Phi^2
+ \frac{\mu}{2}  \omega^{ab} F_{ab}\Phi\\
\phantom{S_{\rm phys}  =}{}
 +\left(\frac{1}{2}\bar \Upsilon^{\dot \alpha} D_5  \Upsilon_{\dot\alpha} + \frac{i}{2} \mu \bar
\Upsilon^{\dot\alpha} \Upsilon_{\dot{\a}}+{\rm h.c.}\right)
+\frac{1}{4} \bar \Xi^{\dot \alpha} \left(- \frac{2i}{\mu} \nabla_{S^2}\right)^2 \Xi_{\dot \alpha}  -
\d_\mu \bar \Xi^{\dot \alpha}\d^\mu \Xi_{\dot \alpha} \\
\phantom{S_{\rm phys}  =}{}
-  \frac{3}{2} \mu^2\bar \Xi^{\dot \alpha}  \Xi_{\dot \alpha}
 +\frac{1}{4}\bar \Lambda \slashed{\d}\Lambda +\frac{1}{4}  \bar g_a \slashed{\d} g^a +
\frac{i}{4}\omega^{ab}\bar G_{ab}\Lambda + \frac{i}{2}\mu \bar \Lambda \Lambda
\Bigg] .
\end{gather*}

\section{Conclusions}\label{conclusions}

In this paper we reviewed our fuzzy $S^2$ construction in terms of bifundamental matrices, originally obtained in the context of the ABJM model in~\cite{Nastase:2009ny,Nastase:2009zu}, focusing on its model-independent mathematical aspects.  We found that this is completely equivalent to the usual adjoint $\SU(2)$ construction, but that it involves fuzzy versions of Killing spinors on the 2-sphere, which we def\/ined.  We described the qualitative dif\/ferences that appear when using the bifundamental $S^2$ to `deconstruct' higher dimensional f\/ield theories. The expansion of the f\/ields involving fuzzy Killing spinors result in an automatic twisting of the former on the sphere. Alternatively, including the Killing spinors in the fuzzy spherical harmonic expansion provides a new approach to the construction of f\/ields on~$S^2$.

We expect that the generality of the construction will lead to it f\/inding a place in numerous applications both in the context of
physical systems involving bifundamental matter,  e.g.\ quiver gauge theories as in~\cite{Maldacena:2009mw}, as well as noncommutative
geometry. We hope to further report on both of these aspects in the future.

\subsection*{Acknowledgements}

It is a pleasure to thank Sanjaye Ramgoolam for many comments, discussions and collaboration in~\cite{Nastase:2009ny}. CP is supported by the STFC grant ST/G000395/1.

\pdfbookmark[1]{References}{ref}
 \LastPageEnding

\end{document}